\documentclass[useAMS,usenatbib]{mn2e}
\usepackage{subfigure}
\usepackage{graphicx}

\title[The environments of radio galaxies and type-2 quasars]{The environments of luminous radio galaxies and type-2 quasars}
\author[C. Ramos Almeida et al.]
{\parbox{\textwidth}{C. Ramos Almeida$^{1,2}$\thanks{E-mail: cra@iac.es},
P.~S.~Bessiere$^{3}$,
C.~N.~Tadhunter$^{3}$,
K.~J.~Inskip$^{4}$,
R.~Morganti$^{5,6}$,
D. Dicken$^{7}$, 
J. I. Gonz\'alez-Serrano$^{8}$ \&
J.~Holt$^{9}$
}\vspace{0.4cm}\\
\parbox{\textwidth}{$^{1}$Instituto de Astrof\' isica de Canarias, Calle V\' ia L\'actea, s/n, E-38205, La Laguna, Tenerife, Spain\\
$^{2}$Departamento de Astrof\' isica, Universidad de La Laguna, E-38205, La Laguna, Tenerife, Spain\\
$^{3}$Department of Physics and Astronomy, University of Sheffield, Sheffield, S3 7RH, UK\\
$^{4}$Max-Planck-Institut f\"ur Astronomie, K\"oningstuhl 17, D-69117 Heidelberg, Germany\\
$^{5}$Netherlands Institute for Radio Astronomy, Postbus 2, 7990 AA Dwingeloo, the Netherlands\\
$^{6}$Kapteyn Astronomical Institute, University of Groningen, Postbus 800, 9700 AV Groningen, the Netherlands\\
$^{7}$Institute d'Astrophysique Spatiale, Centre universitaire d'Orsay, ORSAY CEDEX, France \\
$^{8}$Instituto de F\' isica de Cantabria, CSIC-Universidad de Cantabria, E-39005, Santander, Spain\\
$^{9}$Leiden Observatory, Leiden University, PO Box 9513, 2300 RA Leiden, the Netherlands}}

\begin{document}

\date{}

\pagerange{\pageref{firstpage}--\pageref{lastpage}} \pubyear{2012}

\maketitle

\label{firstpage}

\begin{abstract}
We present the results of a comparison between the environments of 1) a complete sample of 46 southern 
2Jy radio galaxies at intermediate redshifts ($0.05<z<0.7$), 2) a complete sample of 20 radio-quiet type-2 
quasars ($0.3\leq z\leq 0.41$),
and 3) a control sample of 107 quiescent early-type galaxies at $0.2\leq z<0.7$ in the Extended Groth Strip (EGS).
The environments have been quantified using angular clustering amplitudes (B$_{gq}$) derived from deep optical 
imaging data. Based on these comparisons, we discuss the role of the environment in the triggering 
of powerful radio-loud and radio-quiet quasars.
When we compare the B$_{gq}$ distributions of the type-2 quasars and quiescent early-type galaxies, we find
no significant difference between them. This is consistent with the radio-quiet quasar phase being a short-lived but
ubiquitous stage in the formation of all massive early-type galaxies. On the other hand, PRGs are in denser environments 
than the quiescent population, and this difference between distributions of B$_{gq}$ is significant at the 3$\sigma$ level. 
This result supports a physical origin of radio loudness, with high density gas environments favouring the transformation of AGN
power into radio luminosity, or alternatively, affecting the properties of the supermassive black holes themselves.
Finally, focussing on the radio-loud sources only, we find that the clustering of weak-line 
radio galaxies (WLRGs) is higher than the strong-line radio galaxies (SLRGs), constituting a 3$\sigma$ result.
82\% of the 2Jy WLRGs are in clusters, according to our definition (B$_{gq}\ga 400$) versus only 31\% of the SLRGs.
\end{abstract}


\section{Introduction}
\label{intro}

Quasars have long played an important role in the study of galaxy evolution. Initially seen as 
exotic objects, their highly luminous optical, and sometimes also radio, emission led to their 
use as probes of the high redshift universe. More recently, we have seen widespread acceptance 
for the ubiquity of the supermassive black holes that power their active nuclei, and the likelihood
that feedback during the AGN phase may play an important role in moderating galaxy formation and 
evolution. However, we know surprisingly little about how and when quasars are triggered as part 
of the hierarchical growth of galaxies (see \citealt{Alexander12} for a recent review). 

From a theoretical standpoint, simulations of hierarchical galaxy evolution predict that the
periods of black hole growth and nuclear activity are intimately tied to the growth of the host
galaxy \citep{Kauffmann00,diMatteo05,Springel05,Hopkins08a,Hopkins08b,Somerville08}. The tidal torques 
associated with galaxy bars, disc instabilities, galaxy interactions and major
mergers between galaxies are efficient mechanisms to transport the cold gas required to trigger and
feed AGN and star formation to the centre of galaxies. The gas has to lose $\sim$99.9\% of its
angular momentum to travel from the kpc-scale host galaxy down to $\sim$10 pc radius \citep{Jogee06}.



From the observational point of view, imaging studies of samples of luminous, quasar-like
AGN (L$_{bol}>10^{45}~erg~s^{-1}$) have revealed a high incidence of tidal features in their host galaxies 
(\citealt{Heckman86,Hutchings87,Smith89,Canalizo01,Canalizo07,Bennert08}, Ramos Almeida et al. 
2011a, \citealt{Bessiere12}). These tidal features are the result of a past or on-going interaction 
with another galaxy, indicating that galaxy mergers/interactions likely play a role in the triggering of 
powerful AGN. 
Galaxy interactions are one of the most efficient mechanism to transport the cold gas required to trigger and feed
AGN to the center of galaxies \citep{Kauffmann00,Cox06,Cox08,Croton06,diMatteo07}.

In our previous work (\citealt{Bessiere12}, Ramos Almeida et al. 2011a; hereafter \citealt{Ramos11}) 
we studied the optical morphologies of complete samples of 46 southern 2Jy radio galaxies at intermediate redshifts 
($0.05<z<0.7$) and 20 type-2 radio-quiet quasars at $0.3\leq z\leq 0.41$.
We found that the overall majority of the samples (85\% of the PRGs and 75\% of the type-2 quasars) 
show tidal features of relatively high surface brightness. 
In Ramos Almeida et al. (2012; hereafter \citealt{Ramos12}) and \citet{Bessiere12}, we compared the PRG and 
type-2 quasar morphologies with those of a control sample of early-type galaxies matched in redshift, luminosity and 
angular resolution. When we considered the same surface brightness limits, the fraction of disturbed 
morphologies in the quiescent population was considerably smaller than in the PRGs and type-2 quasars. 
This supports a scenario in which radio-loud and radio-quiet quasars represent a fleeting active phase 
of a subset of the elliptical galaxies that have recently undergone mergers/interactions. 

Another factor that can have an influence on how AGN are triggered is the environment. Previous
studies have shown that intermediate to low-redshift radio-quiet quasars reside in groups rather than in rich
clusters (e.g. \citealt{Fisher96,Bahcall97,McLure01}). More recently, \citet{Serber06} studied the 
environment of $\sim$2,000 quasars at redshift $z\leq0.4$ on different scales, using data from the SDSS survey. 
The latter authors claim that, on scales of $\sim$1 Mpc, the environments of quasars are not significantly different 
from those of quiescent L$_*$ galaxies. On smaller scales, specifically the inner $\sim$100 kpc, they found
a dependence of quasar environment on luminosity. The more luminous
the quasars, the richer the environments. This enhanced galaxy density on a $\sim$100 kpc scale is consistent with
luminous quasars residing in galaxy groups --just the type of environment that is likely to favour galaxy mergers
and interactions.

The case of radio-loud AGN may be different. Past investigations have shown
mixed results. On the one hand, several works have found a difference
between the environment of radio-loud and radio-quiet quasars.
Low--to--intermediate redshift radio galaxies are generally found in Abell 0--1 clusters, whereas radio-quiet
quasars normally reside in groups \citep{Yee84,Yee87,Ellingson91,Wold00,Wold01,Best05,Kauffmann08,Falder10}. 
This difference in the clustering of radio-loud and radio-quiet AGN could 
imply that the environment has an influence in the radio luminosity of active galaxies. 
On the other hand, in a study of the environments of a sample of 44 PRGs and luminous quasars at $z\sim0.2$, 
\citet{McLure01} did not find a significant difference in the clustering of the two groups. They claimed that
both inhabit environments that are compatible with Abell 0 class. 

Studies of the environment of AGN may also help us to distinguish between models that seek to explain
the relationship between different classes of AGN. For example, it has been proposed that luminous AGN could cycle 
between radio-loud and radio-quiet phases within a single quasar triggering event (see e.g. \citealt{Nipoti05}).
If radio-quiet and radio-loud sources are the same object going through a different phase, then we should find
similar environments for the two of them on the same scales.

In \citealt{Ramos11} we found that galaxy interactions likely play a key role in the triggering of AGN/jet activity, 
especially in the case of strong-line radio galaxies (SLRGs)\footnote{SLRGs 
comprise narrow- and broad-line radio galaxies and quasars, i.e.
they are radio galaxies with strong and high equivalent width emission lines.}, of which 94\% appear disturbed.
However, a subset of the 2Jy sample presents optical 
morphologies and emission-line kinematics that do not support the idea of the AGN triggering via mergers. 
These include some central cluster galaxies surrounded by massive haloes of hot gas 
\citep{Tadhunter89,Baum92}. In such cases, the infall of cold 
gas condensing from the X-ray haloes in cooling flows has been suggested as a triggering mechanism 
\citep{Tadhunter89,Baum92,Bremer97,Edge99,Edge10}. 
Moreover, it has been shown that the direct accretion of hot gas from the X-ray haloes of galaxies 
is a plausible mechanism for fuelling radio galaxies that lack strong emission lines, namely the 
weak-line radio galaxies (WLRGs; \citealt{Allen06,Best06,Hardcastle07,Balmaverde08,Buttiglione10})\footnote{WLRGs 
have optical spectra dominated by the stellar continua of the host galaxies and small emission line 
equivalent widths (EW$_{[O III]}<10$ \AA; \citealt{Tadhunter98}).}. It turns out that only 27\% of the WLRGs 
in the 2Jy sample show clear evidence for tidal features, supporting the hypothesis of, at least some of them, 
being triggered by a different mechanism than the SLRGs (see also \citealt{Best05,Sabater13}). 

Considering the radio morphological classification of PRGs, the environment of 
low redshift Fanaroff-Riley I (FRI) PRGs appears to be richer than their Fanaroff-Riley II (FRII) 
counterparts \citep{Prestage88,Prestage89,Zirbel97,Gendre13}.  
The majority of FRII galaxies in the 2Jy sample are classified as SLRGs in the optical, 
with a minority showing WLRG spectra. On the other hand, all FRI galaxies in the 2Jy sample are WLRGs 
according to their optical spectra. If the 2Jy WLRGs/FRIs are found in denser environments than SLRGs/FRIIs, 
that would support the hypothesis that AGN are either fuelled by warm gas condensing out of the 
hot X-ray haloes of clusters \citep{Tadhunter89,Baum92,McDonald11,McDonald12}, or by direct accretion of hot gas 
\citep{Best06,Hardcastle07}.

This is the fourth in a series of papers based on the analysis of the optical morphologies of complete samples 
of PRGs, type-2 quasars, 
and quiescent early-type galaxies (\citealt{Ramos11}, \citealt{Ramos12} and \citealt{Bessiere12}; see Table \ref{samples}). 
Here we study the influence of the environment on the triggering and fuelling of the AGN.
In Section 2 we describe the different samples, the observations employed and how the catalogs were constructed. 
In Section 3 we present the 
results on the galaxy enviroments. The comparison between the environments of PRGs, type-2 quasars and quiescent 
elliptical galaxies
is discussed in Section 4, and the main conclusions from this work are summarized in Section 5. 
Throughout this paper we 
assume a cosmology with H$_0$ = 70 km s$^{-1}$ Mpc$^{-1}$, $\Omega_m$ = 0.27, 
and $\Omega_{\Lambda}$ =0.73.

\begin{table}
\centering
\footnotesize
\begin{tabular}{lccr}
\hline
\hline
Sample                     & Sources       & Redshift 	      & Interactions  \\
		           &		   &		      & (per cent)  \\
\hline
2Jy PRGs (SLRGs)	   &  46 (35)      & $0.05<z<0.7$     		& 85 (94) (a)  \\
EGS early-types    	   & 107 	   & $0.2\leq z<0.7$  		& 53 	(a)    \\
Type-2 quasars 		   &  20 	   & $0.3\leq z\leq 0.41$     	& 75 	(b)    \\
EGS* early-types    	   &  51 	   & $0.3\leq z\leq 0.41$ 	& 57 	(b)    \\
\hline		     			      		 
\end{tabular}						 
\caption{Galaxy samples considered in this work. The fraction of signs of interaction was calculated considering 
features with $\mu_V\leq$ 25.5~mag~arcsec$^{-1}$. Values between parenthesis correspond to SLRGs only. 
EGS and EGS* are the control samples for the PRGs and type-2 quasars respectively.
References: (a) \citealt{Ramos12}; (b) \citet{Bessiere12}.}
\label{samples}
\end{table}

\section{Sample Selection, observations and catalogs}
\label{selection}

\subsection{The 2Jy sample of PRGs} 
\label{PRG}

The objects studied in~\citealt{Ramos11} comprise all PRGs from the \citet{Tadhunter93} 
sample of 2Jy radio galaxies with S$_{2.7GHz}\ge$ 
2.0 Jy, steep radio spectra $\alpha_{2.7}^{4.8} > 0.5~(F_{\nu}\propto\nu^{-\alpha})$, 
declinations $\delta<+10\degr$ and redshifts $0.05<z<0.7$ (see Table 1 in \citealt{Ramos11}). 
It is itself a subset of the \citet{Wall85} complete sample of 2Jy radio sources.
The z $>$ 0.05 limit ensures that the radio galaxies are genuinely powerful sources, while the z $<$ 0.7 limit ensures
that sources are sufficiently nearby for detailed morphological studies.

In terms of the optical classification, based on both previous optical spectra \citep{Tadhunter98}
and on optical appearance \citep{Wall85}, the sample comprises 24\% WLRGs, 43\% Narrow-Line Radio Galaxies (NLRGs), 
and 33\% Broad-Line Radio Galaxies and quasars (BLRGs and QSOs). 

Considering the radio morphologies, FRII sources constitute the majority of the sample (72\%),
13\% are FRI, and the remaining 15\% correspond to compact, steep-spectrum (CSS) or 
Gigahertz-peaked spectrum (GPS) sources (see Table 1 in \citealt{Ramos11}).

Our sample of 46 PRGs was
imaged with the Gemini Multi-Object Spectrograph South (GMOS-S) on the 8.1-m Gemini South telescope at Cerro Pach\'on
under good seeing conditions (median seeing full width at half maximum (FWHM) of 0.8\arcsec, ranging from 0.4\arcsec~to 1.1\arcsec).
The seeing values were measured individually for each of the 46 GMOS-S images, using foreground stars. 
The GMOS-S detector \citep{Hook04} comprises three adjacent CCDs, giving a field-of-view (FOV) of 5.5$\times$5.5 
arcmin$^2$, with a pixel 
size of 0.146\arcsec. The morphological features reported in \citealt{Ramos11} have surface brightness 
within the range $21\leq\mu_V\leq26$~mag~arcsec$^{-2}$, with a median value of $\mu_V$=23.6~mag~arcsec$^{-2}$. 

With the exception of the source PKS 2250-41, all the galaxies with z$\leq$0.4 were observed 
in the r'-band filter ($\lambda_{eff}$=6300 \AA, $\Delta\lambda$=1360 \AA), while those with $z>0.4$ 
were observed in the i'-band ($\lambda_{eff}$=7800 \AA, $\Delta\lambda$=1440 \AA), to cover the 
typical rest-frame wavelength range 4500-6000 \AA.
See \citealt{Ramos11} for a more detailed description of the GMOS-S observations.

\subsection{The type-2 quasar sample} 
\label{type2}

In \citet{Bessiere12} we performed the same morphological analysis as in \citealt{Ramos11}, but for a sample
of 20 type-2 quasars selected from \citet{Zakamska03}, with
right ascensions (RAs) $23^h<RA<10^h$, declinations $\delta<20^{\circ}$, 
redshifts between 0.3 and 0.41 and [O III] luminosities larger than 10$^{8.5}L_{\sun}$ (see Table 1 in 
\citealt{Bessiere12}). The [O III] luminosity limit was chosen to ensure the quasar nature of the sources. The 
full sample of 20 objects is complete and unbiased in terms of host galaxy properties. 

Deep optical imaging data for the 20 objects were obtained using GMOS-S and exactly the same
instrumental configuration as for the 2Jy sample (see Section \ref{PRG}). The observations were carried 
out in queue mode between 
2009 August and 2011 September in good seeing conditions, with a median value of 
FWHM = 0.8\arcsec, ranging between 0.5\arcsec~and 1.1\arcsec. Due to the redshifts of the type-2 quasars, observations 
were done using the r'-band filter only. The surface brightnesses of the tidal features detected 
are within the range $21\leq\mu_V\leq25$~mag~arcsec$^{-2}$, with a median value of $\mu_V$=23.4~mag~arcsec$^{-2}$.
A summary of the observations can be found in Table 2 in \citet{Bessiere12}.

As well as the main science target fields, one offset field ($\sim$20 arcmin offset) 
was observed after each radio galaxy and type-2 quasar observation, in order to better quantify the background 
galaxy population of the host galaxies. The offset field observations were taken immediately after 
the science targets\footnote{The only exceptions are PKS 1602+01 and PKS 1814-63, whose corresponding offset fields
were observed on different nights, but under similar seeing conditions.} and with the same or longer exposure times 
(from 800 to 1500 s).
Unfortunately, we do not have offset field observations for three
of the type-2 quasars, namely J0025-10, J0159+14 and J0142+14. Therefore, we have 46 offset fields for the PRGs 
and 17 for the type-2 quasars (i.e. 52 offset fields in total in the r'-band and 11 in the i'-band).

\subsection{Control sample of quiescent early-type galaxies}
\label{quiescent}

In \citealt{Ramos11} we analysed the optical morphologies of the 2Jy sample of PRGs and found a large fraction 
(85\%) of disturbed galaxy hosts. In order to study the importance of galaxy interactions in the AGN triggering
phenomena, we developed a control sample of non-active (quiescent) galaxies to classify their morphologies in 
exactly the same way. 
Since radio galaxies are almost invariably associated with elliptical hosts (see e.g. \citealt{Heckman86} and 
\citealt{Dunlop03}), we searched in the literature for 
samples of early-type galaxies with similar masses and redshifts as the 2Jy PRGs. In addition, we required
similar angular resolutions
and depths to probe the same spatial scales and surface brightness levels. 
After considering all these factors, we finally selected control samples of early-type 
galaxies in two redshift ranges which best matched the 2Jy sample host galaxies: the Observations of 
Bright Ellipticals at Yale (OBEY) survey (55 elliptical galaxies with redshifts $z<0.2$) and the Extended 
Groth Strip (EGS) sample (107 early-type galaxies with redshifts $0.2\leq z<0.7$). For the type-2 quasars, 
we selected a separate control sample from the EGS to match the redshift and absolute magnitude ranges of this 
sample \citep{Bessiere12}. 

The goal of this paper is to quantify the environments of PRGs, type-2 quasars and quiescent galaxies to 
try to understand the role that it plays, if any, in triggering AGN. Unfortunately, the OBEY survey images that we 
used in \citealt{Ramos12} to classify the galaxy morphologies 
are not suitable for the study of the environment, 
because of the limited FOV of Y4KCam at CTIO and the low redshift of the sources. Therefore, in the 
following we will refer only to the EGS galaxies 
as the control sample (see Table \ref{samples}).

We selected our EGS control sample ($\alpha$ = 14$^{h}~17^{m}$, 
$\delta$ = +52$\degr~30^{\prime}$) using the {\it Rainbow Cosmological 
Surveys database}\footnote{https://rainbowx.fis.ucm.es/Rainbow$_{-}$Database},
which is a compilation of photometric and spectroscopic data, jointly with value-added products such as photometric 
redshifts, stellar masses, star formation rates, and synthetic rest-frame magnitudes, for several deep cosmological fields 
\citep{Perez08,Barro09,Barro11}. We used the publicly available broadband images of the EGS obtained with the 
Subaru Prime Focus Camera (Suprime-Cam; \citealt{Miyazaki02}), taken as part of the Subaru Suprime-Cam Weak-Lensing Survey \citep{Miyazaki07}. 
Four pointings of 30 min exposure time each in the 
R$_c$ filter were necessary to cover the entire EGS to a limiting AB magnitude of R$_c\sim$26 (\citealt{Barro11}; see also
Appendix \ref{appendixA}). 
The detector of Suprime-Cam is a mosaic of ten 2048$\times$4096 CCDs located at the prime focus of Subaru Telescope, and   
it covers a 34$\times$27 arcmin$^2$ FOV with a pixel scale of 0.202\arcsec. In \citealt{Ramos12}
we measured a median surface brightness of $\mu_V$=24.2~mag~arcsec$^{-2}$ for the tidal features detected, 
and a surface brightness range $22\leq\mu_V\leq26$~mag~arcsec$^{-2}$.
The seeing of the 4 images ranges from FWHM = 0.65\arcsec to 0.75\arcsec. 
Thus, the data are comparable in depth and resolution to the GMOS-S images employed in the study 
of PRGs and type-2 quasars. For further details on the observations of the EGS, we refer the reader
to \citet{Zhao09}.

We selected all the galaxies in the EGS to fall in the same redshift and absolute magnitude ranges as the PRGs 
at $z\geq0.2$ in \citealt{Ramos11} ($0.2\leq z<0.7$ and $-22.2\leq M_B \leq -20.6$ mag respectively). 
From this first selection we discarded the sources in the EGS detected in X-rays
(i.e. possible AGN) and foreground stars. The stars were automatically identified based on a combination of 
several criteria 
including their morphology (stellarity index) and their optical/NIR colours (see \citealt{Perez08} and 
\citealt{Barro11} for details on the star-galaxy separation criteria).

In order to identify early-type galaxies, we imposed a colour selection criterion: initially we selected
all the sources with rest-frame colours (M$_u$-M$_g$) $>$ 1.5, typical of galaxies located in the red sequence 
in the colour-magnitude diagram 
\citep{Blanton06}.
After applying the colour selection, 
we made a first visual classification of the sources into three groups: elliptical galaxies (E), 
possible disks (PD), and disks (D). 
We then discarded all the galaxies that appeared as clear disks and kept the elliptical galaxies and 
possible disks in the sample.
The latter might include disturbed ellipticals that look more disk-like, or S0/early-type spirals. 
After considering all these criteria, we have a control sample of 107 red early-type galaxies in the EGS 
matched in redshift and absolute magnitude to the 2Jy sample (see Table 2 and Figures 2 and 3 in \citealt{Ramos12}). 


For comparison with the type-2 quasar host galaxies, we repeated the same procedure as for the PRGs, but
adjusting the ranges of absolute magnitude and redshift to be the same as the type-2 quasar sample. 
Thus, we selected galaxies in the EGS sample in the redshift range $0.3\leq z\leq 0.41$, with 
absolute magnitudes $-22.1\leq M_B \leq -20.3$ mag and rest-frame colours (M$_u$-M$_g$) $>$ 1.5.
This leaves us with a comparison sample of 51 quiescent early-type galaxies. In the following, 
we will refer to the control samples of the PRGs and type-2 quasars as EGS and EGS*, respectively (see Table \ref{samples}).  
See \citealt{Ramos12} and \citet{Bessiere12} for further details on the control sample selection.

\subsection{Galaxy catalogs}
\label{catalogs}

Our aim is to quantify the richness of the environments of PRGs, type-2 quasars and control sample galaxies. 
Since we do not have spectroscopic redshifts for all the sources detected in the galaxy fields, 
we need a reliable estimate of the number of galaxies in the vicinity of the targets. 
Thus, we used the spatial cross-correlation function to characterise our sources environments. 
This technique has the advantage of requiring just one wide-field image in a single filter,
and it is based on a statistical approach, consisting of the normalization of the surface densities using the field galaxy 
luminosity function. 

The first step of this analysis involved generating the galaxy catalogs. For that purpose we used the Graphical Astronomy and
Image Analysis tool (GAIA), which has an interactive toolbox facility that uses the program EXTRACTOR and 
Source Extractor (SExtractor, v.2.5.0; \citealt{Bertin96}). SExtractor automatically detects and parameterises all the 
sources in an input image with fluxes above a threshold level defined by the user. These objects are then identified 
by elliptical contours over the image and are available for interactive inspection. The resulting measurements, 
including magnitudes computed using different standard methods, are then recorded in catalogues. 
The SExtractor input parameters employed in the construction of the galaxy catalogs for the fields of PRGs, 
type-2 quasars, control sample galaxies and corresponding offset fields are reported in Table \ref{SExtractor}.

\begin{table*}
\centering
\footnotesize
\begin{tabular}{llcc}
\hline
\hline
Parameter & Description & GMOS-S  & Suprime-Cam \\
\hline
DETECT$_-$MINAREA	& Min number of pixels above threshold	        & 5   	    				&	5			\\ 	
DETECT$_-$THRESH	& Detection threshold				& 5					&	7			\\
ANALYSIS$_-$THRESH 	& Surface brightness threshold  		& 1.5					&	1.5			\\
DEBLEND$_-$NTHRESH 	& Number of deblending sub-thresholds		& 32					&	32			\\
DEBLEND$_-$MINCONT 	& Min contrast parameter for deblending 	& 0.0001				&	0.0001			\\
CLEAN$_-$PARAM     	& Efficiency of cleaning		 	& 5					&	5			\\
MAG$_-$ZEROPOINT	& Magnitudes zeropoint offset           	& (r') 28.32+2.5Log(t)-0.10(AIRM-1)	&  (Rc) 31.85,31.82,31.79,31.86	\\
			&						& (i') 27.92+2.5Log(t)-0.08(AIRM-1)	&				\\
PIXEL$_-$SCALE     	& Pixel size in arcsec				& 0.146					&	0.202			\\
GAIN            	& In e$^-$/ADU					& 5.0					&	2.5			\\
BACK$_-$SIZE       	& Size of the background mesh      		& 100,125,150,175,200*			&	175			\\
BACK$_-$FILTERSIZE 	& Size of the background-filtering mask		& 3					&	3			\\	   
\hline		     			      		 
\end{tabular}						 
\caption{SExtractor input parameters. (*) Chosen to match MAG$_-$APER of the PRGs and type-2 quasars with the values reported in \citealt{Ramos11} 
and \citet{Bessiere12}, as calibration.}
\label{SExtractor}
\end{table*}

The parameter choice was done in two steps. First, we followed the indications provided in the SExtractor 
manual and the values chosen in similar studies (e.g. \citealt{Ryan10}). Second, we refined our parameter 
choice by forcing the aperture magnitudes in the catalogs (MAG$_-$APER) to match those reported in 
\citealt{Ramos11} and \citet{Bessiere12} for the PRGs and type-2 quasars respectively (see Table \ref{SExtractor}). 
Magnitude zeropoints were individually calculated for the PRGs (in the r'- and i'-bands) and type-2 quasars 
(r'-band) using corresponding exposure time and airmass. In the case of the EGS sample, each of the
four Subaru fields has a different zeropoint (see Table \ref{SExtractor}). Thus, we produced individual galaxy catalogs for 
each PRG, type-2 quasar and offset field, plus large catalogues for each of the four Subaru fields. 

Among the different instrumental magnitudes provided by SExtractor, we chose the automatic aperture
magnitudes (MAG$_-$AUTO), which are precise estimates of the total galaxy magnitudes. This routine
is based on the \citet{Kron80} ``first moment'' algorithm\footnote{For further details on the automatic aperture
magnitude determination we refer the reader to the SExtractor manual:
http://www.astromatic.net/software/sextractor.}. 
To discriminate stars from galaxies we used the detection parameter CLASS$_-$STAR, which is equal 
to 0 when the source is a galaxy, and 1 if it is a star. Values in between have a more ambiguous 
interpretation, but we can assume that the closer CLASS$_-$STAR to 1, the more likely the classification
of the object as a star. When the sources contained in the catalogs are bright, the distribution 
of CLASS$_-$STAR values is roughly bimodal, and becomes less accurate for fainter sources \citep{Ryan10}.
Ground-based studies by \citet{Fadda04} and \citet{Ryan10} found CLASS$_-$STAR$\leq0.85$ to be a good criterion
to select extended sources when the objects are brighter than R = 23 mag. In addition, to get rid of 
possible intruder stars in our galaxy catalogs, we restricted the range of apparent magnitudes in the 
final catalogs (see Section \ref{counting}).  

Finally, to discard sources close to image boundaries, or with saturated and/or corrupted pixels, 
we used the detection parameter FLAG. Sources with FLAG$>$4 are removed from catalogs. Objects
with neighbors and/or bad pixels (FLAG=1), originally blended with another object (FLAG=2) 
or with a combination of the two (FLAG=3) are included in the catalogs in addition to the non-compromised objects
(FLAG=0). 
Once a blended object is extracted, the connected pixels pass through a 
filter that splits them into overlapping components. This normally happens
if the field is crowded and/or if the detection threshold is low.

\subsection{Galaxy counting}
\label{counting}

In the same manner as in \citet{McLure01}, we counted galaxies around our PRGs, type-2 quasars and control sample
galaxies which satisfy the following two criteria:

\begin{enumerate}
\renewcommand{\theenumi}{(\arabic{enumi})}
\item the galaxies are at a projected distance from the central source less than the counting radius, which 
is defined by the object with the lowest redshift among the three samples considered. In our case it is 
the radio galaxy PKS 0620-52 (z=0.051). For this source redshift, the distance between the radio galaxy
and the edge of the GMOS-S field corresponds to 170 kpc in the chosen cosmology. 
Therefore, we employed this projected radius for counting galaxies around all the targets considered in this paper. 
For the GMOS-S and Subaru offset fields, we first counted 
all galaxies within a circle of radius equal to half of the size of the CCD field (r$_{im}$). Second, we divided that number
of galaxies by the area of that circle ($\pi r_{im}^2$), and finally, multiplied by the area of a circle 
of 170 kpc radius ($\pi r_{170 kpc}^2$).

\[
N = N_{im} \times \frac{r_{170 kpc}^2}{r_{im}^2}
\]

Although this projected radius is among the smallest considered in environment studies (e.g. \citealt{Serber06}), 
it should be sufficient for studying the clustering around AGN. The reason is the slope of the two-point correlation
function that we assumed ($\gamma$=1.77; \citealt{Groth77}). This slope allows a reliable study of the 
clustering around AGN even when restricted to scales of 100-200 parsecs \citep{McLure01}.

\item The galaxies included in N$_t$ (total number of galaxies within a r$_{170 kpc}$ radius, excluding the target) 
and N$_b$ (number of background galaxies within the same radius) are required to have similar magnitudes to a generic galaxy at the 
redshift of the target. We adopted the same criterion as in \citet{McLure01}: $(m_*-1)\leq m\leq (m_*+2)$. 
In the case of a galaxy cluster, this range will include the galaxies containing the majority of the cluster mass. 

Therefore, we first calculated the theoretical value of M$^*_B$ at the redshift of all our targets 
using the evolution with redshift
of the Schechter function parameters given in \citet{Faber07} for the ``All galaxy sample''. This sample 
includes galaxies with redshifts z$\leq$1 from DEEP2 and COMBO-17. The next step is to transform those
absolute magnitudes into apparent ones (m$_*$) in the r', i' and Rc bands, to make them comparable to our
targets magnitudes. To do that, we assumed 
colors of Sbc galaxies, which are intermediate between those of early and late-type galaxies. 
We also need to remove the corresponding reddening and K-corrections performed in \citealt{Ramos11},\citealt{Ramos12} and \citet{Bessiere12}, 
to obtain apparent magnitudes comparable to those in our galaxy catalogs. 
For the GMOS-S offset fields we used values of the reddening measured in center of each field 
from the NASA/IPAC Infrared Science Archive (IRSA). 
Finally, for each target, 
we used the calculated m$_*$ value --which in general is dimmer than the PRGs and type-2 quasars, and 
similar to the control sample galaxies-- to count the galaxies included in the interval 
[m$_*$-1, m$_*$+2] in both the target and offset fields. Since we are counting galaxies in images taken 
with different instruments, exposure times and seeing conditions, it is
necessary to assess whether those data are deep enough to count galaxies down to the dimmest limit of 
the magnitude interval (m$_*$+2). This analysis is presented in Appendix \ref{appendixA}. 
\end{enumerate}

\subsection{Spatial clustering amplitude}
\label{cluster}

Our aim is to determine spatial clustering amplitudes (B$_{gq}$; \citealt{Longair79}) for all the individual objects
in our complete PRG, type-2 quasar and control galaxy samples. This is a widely used technique that allows direct 
comparison with previous studies \citep{Longair79,Prestage88,Ellingson91,Hill91,Yee99,McLure01,Ryan10}.

First, we need to determine the angular correlation function
\[
n(\sigma)\delta\Omega = N_g[1+w(\sigma)]\delta\Omega,
\qquad w(\sigma) = A_{gq}\sigma^{1-\gamma}.
\]
A$_{gq}$ represents the excess in the number of galaxies around the target as compared with the 
predicted number of background galaxies per unit area, N$_g$. 
\[
A_{gq} = \left[ \frac{N_t}{N_b}-1 \right]\left( \frac{3-\gamma}{2} \right)(\theta_{170kpc})^{\gamma-1}. 
\]
N$_t$ is the total number of galaxies within the $\theta_{170kpc}$ radius (in radians) excluding 
the target (i.e. the PRG, type-2 quasar of control sample galaxy). N$_b$ is the number of background 
galaxies within the same radius, calculated as described in Section \ref{counting}. Finally, 
$\gamma$ is the slope of the two-point correlation function that we have to assume to calculate
the spatial clustering amplitude of the target. Here we consider $\gamma$=1.77, which 
is the slope that better describes the clustering of galaxies around
AGN \citep{Groth77,McLure01}.

To compare the clustering around targets covering a redshift range, we need to de-project the angular
correlation function into its spatial equivalent:
\[
n(r)\delta V = \rho_g[1+\epsilon(r)]\delta V,
\qquad \epsilon(r) = B_{gq}r^{-\gamma}.
\]
By assuming that galaxy clustering is spherically symmetric around the target \citep{Longair79}, we 
can calculate B$_{gq}$ as
\[
B_{gq} = \frac{A_{gq}N_g}{I_{\gamma}\phi(z)}\left( \frac{d}{1+z} \right)^{\gamma-3}. 
\]
The angular-size distance to the target is $d$, and I$_{\gamma}$ = 3.78 for a field-galaxy value of $\gamma$=1.77
\citep{Groth77}. $\phi(z)$ is the integrated luminosity function, above the luminosity limit, at the redshift of the target.
The adopted Schechter function parameters in the different redshift bins and photometric bands considered in this
work are reported in Table \ref{Schechter}. A comparison between predicted and measured background galaxy counts is shown 
in Appendix \ref{appendixB}.

%
%

\begin{table}
\centering
\footnotesize
\begin{tabular}{ccccc}
\hline
\hline
Redshift & M$^*_{r'}$ & M$^*_{i'}$ & M$^*_{Rc}$ & $\phi^*$         \\
bin      & (mag)      & (mag)      & (mag)      & (Gal~Mpc$^{-3}$) \\  
\hline
0.0--0.2 & -21.43 & -21.76 & -21.66 & 0.0038   \\
0.2--0.4 & -22.08 & -22.47 & -22.32 & 0.0037   \\
0.4--0.6 & -22.77 & -23.27 & -23.05 & 0.0035   \\
0.6--0.8 & -22.62 & -23.59 & -23.02 & 0.0033   \\
0.8--1.0 & -22.87 & -23.84 & -23.27 & 0.0031   \\
\hline		     			      		 
\end{tabular}						 
\caption{Schechter parameters in the redshift bins 
and photometric bands considered in this work. Parameters 
were obtained from the ``All Galaxy Samples'' fits in \citet{Faber07}. We considered
$\alpha$=-1.3 in all redshift bins. For the low-redshift 
range, \citet{Faber07} reported values ranging between $\alpha$=-0.9 and -1.2, but as discussed by the latter
authors, the corrections needed to use $\alpha$=-1.3 instead are small and 
can be ignored.}
\label{Schechter}
\end{table}						 


For each PRG and type-2 quasar we have obtained B$_{gq}$ using two different 
approaches: first, using the individual dedicated offset field to work out the number of background galaxies (N$_b$). 
Second, using all the 
GMOS-S offset fields observed in the same filter as the target (either r' or i'-band) to obtain the average\footnote{Here and 
throughout all the text, we refer to the average (av) as the arithmetic mean of a sample and/or distribution.} and median 
number of background galaxies ($N_b^{av}$ and $N_b^{med}$). A few offset fields are very crowded, and they are 
significant outliers in terms of their N$_b$ values. In order to avoid the effect that this might have on the 
individual N$_b^{av}$ values, we discarded offset fields with N$_b>\bar{N_b}+\sqrt{\bar{N_b}}$ and then recalculated
the individual $N_b^{av}$, $N_b^{med}$ and $\sigma$ values reported in Tables \ref{Bgq_individual} and \ref{Bgq_individual2}. 
Thus, depending on the redshift of each source, and consequently on 
the counting radius (170 kpc), we used between 40 and 49 offset fields in the r'-band, and between 8 and 10 
in the i'-band to calculate the individual $N_b^{av}$ and $N_b^{med}$ values. 
In Tables \ref{Bgq_individual} and \ref{Bgq_individual2} we report dedicated, average and median values of 
N$_b$ and B$_{gq}$ for the PRGs and type-2 quasars respectively.

For the control sample galaxies, we considered the Subaru field 
in which each target is included as the dedicated offset field, and the four Subaru fields
to work out N$_b^{av}$ and N$_b^{med}$. Individual values are reported in Tables 
\ref{Bgq_control_individual} and \ref{Bgq_control_individual2} in Appendix \ref{appendixC}.

The method employed here aims to quantify the excess of galaxies around the targets as
compared with the number of background galaxies. Therefore, it appears more reliable to use 
N$_b^{av}$ and N$_b^{med}$, which have been calculated using all the available offset fields in a given filter.
However, for low-redshift targets, we found no background galaxies within the counting radius and magnitude
range in the majority of the offset fields, leading to N$_b^{med}$=0 (see Table \ref{Bgq_individual}). 
The same happened with the dedicated N$_b$ values of the PRGs PKS 0625-35, PKS 1814-63, PKS 2356-61 and 
PKS 1599+02. Overall, we consider that B$_{gq}^{av}$ is the most robust measurement of the
environments of PRGs, type-2 quasars and control sample galaxies. We calculated individual errors 
using the same prescription as in \citet{Yee99} and \citet{McLure01}: 

\[
\frac{\Delta B_{gq}}{B_{gq}} = \frac{[(N_t-B_b)+1.3^2 N_b]^{\frac{1}{2}}}{(N_t-N_b)}.
\]

\begin{table*}
\centering
\begin{tabular}{lcccccccccccl}
\hline
\hline
PKS ID   &   z    & Optical & Radio & N$_t$ & N$_b$ &  B$_{gq}$ &  N$_b^{av}$	& $\sigma$ &   B$_{gq}^{av}\pm\Delta B_{gq}^{av}$ & N$_b^{med}$  &  B$_{gq}^{med}$ & Morphology \\
 (1)     &  (2)   & (3)     & (4)   & (5)   & (6)   &  (7)	&  (8)  	& (9)	  &   (10)	     & (11)	    &  (12)	      & (13) \\  
\hline
0620-52  & 0.051  & WLRG & FRI   &   15  & 2.02  &   880  &  0.27 & 0.45&   999$\pm$264  & 0.00  & \dots &  \dots	   \\ 
0625-53  & 0.054  & WLRG & FRII  &   16  & 0.91  &   1025 &  0.25 & 0.41&  1070$\pm$273  & 0.00  & \dots &  B		   \\	   
0915-11  & 0.055  & WLRG & FRI   &   12  & 1.72  &   698  &  0.24 & 0.39&   798$\pm$237  & 0.00  & \dots &  D		   \\ 
0625-35  & 0.055  & WLRG & FRI   &   8   & 0.00  &  \dots &  0.27 & 0.41&   526$\pm$195  & 0.00  & \dots &  J		   \\ 
2221-02  & 0.056  & BLRG & FRII  &   1   & 1.80  &   -54  &  0.27 & 0.42&    50$\pm$74   & 0.00  & \dots &  F,S 	   \\ 
1949+02  & 0.059  & NLRG & FRII  &   5   & 3.21  &   123  &  0.47 & 0.61&   310$\pm$158  & 0.00  & \dots &  S,D 	   \\ 
1954-55  & 0.058  & WLRG & FRI   &   9   & 3.25  &   393  &  0.51 & 0.63&   581$\pm$209  & 0.00  & \dots &  \dots	   \\ 
1814-63  & 0.065  & NLRG & CSS   &   1   & 0.00  & \dots  &  0.70 & 0.79&    21$\pm$83   & 0.66  & 23	 &  2I,D	   \\ 
0349-27  & 0.066  & NLRG & FRII  &   4   & 1.94  &   143  &  0.60 & 0.72&   235$\pm$145  & 0.00  & \dots &  2B,[S]	   \\ 
0034-01  & 0.073  & WLRG & FRII  &   2   & 1.61  &    27  &  0.67 & 0.72&    93$\pm$110  & 0.54  & 102   &  J		   \\ 
0945+07  & 0.086  & BLRG & FRII  &   2   & 1.15  &    60  &  0.79 & 0.69&    86$\pm$113  & 0.77  & 87	 &  S		   \\ 
0404+03  & 0.089  & NLRG & FRII  &   2   & 1.11  &    63  &  0.81 & 0.69&    85$\pm$114  & 0.74  & 90	 &  [S] 	   \\ 
2356-61  & 0.096  & NLRG & FRII  &   7   & 0.00  &  \dots &  0.81 & 0.64&   447$\pm$198  & 0.64  & 459   &  2S,F,I	   \\ 
1733-56  & 0.098  & BLRG & FRII  &   1   & 5.83  &   -349 &  0.89 & 0.69&     8$\pm$90   & 0.61  & 28	 &  2T,2I,2S,[D]   \\ 
1559+02  & 0.105  & NLRG & FRII  &   8   & 0.00  &  \dots &  0.94 & 0.68&   515$\pm$215  & 0.82  & 524   &  2S,D,[2N]	   \\ 
0806-10  & 0.109  & NLRG & FRII  &   9   & 1.01  &   586  &  0.93 & 0.65&   591$\pm$228  & 0.76  & 604   &  F,2S	   \\ 
1839-48  & 0.111  & WLRG & FRI   &   23  & 0.50  &   1657 &  0.97 & 0.66&  1622$\pm$358  & 0.75  & 1638  &  2N,S,[T]	   \\ 
0043-42  & 0.116  & WLRG & FRII  &   4   & 0.46  &   262  &  1.00 & 0.65&   223$\pm$161  & 0.70  & 245   &  [2N],[B]	   \\ 
0213-13  & 0.147  & NLRG & FRII  &   2   & 1.08  &    71  &  1.25 & 0.60&    58$\pm$131  & 1.08  & 71	 &  2S,[T]	   \\ 
0442-28  & 0.147  & NLRG & FRII  &   7   & 0.77  &   482  &  1.33 & 0.69&   438$\pm$218  & 1.23  & 446   &  S		   \\ 
2211-17  & 0.153  & WLRG & FRII  &   19  & 3.19  &   1233 &  1.32 & 0.67&  1379$\pm$348  & 1.16  & 1391  &  D,[F]	   \\ 
1648+05  & 0.154  & WLRG & FRI?  &   8   & 2.55  &   425  &  1.33 & 0.68&   520$\pm$233  & 1.13  & 536   &  D		   \\ 
1934-63  & 0.181  & NLRG & GPS   &   3   & 6.20  &   -259 &  1.53 & 0.68&   119$\pm$163  & 1.41  & 128   &  2N,2T	   \\ 
0038+09  & 0.188  & BLRG & FRII  &   2   & 1.97  &     3  &  1.55 & 0.67&    37$\pm$144  & 1.45  & 45	 &  T		   \\ 
2135-14  & 0.200  & QSO  & FRII  &   6   & 1.40  &   381  &  1.63 & 0.67&   362$\pm$221  & 1.49  & 373   &  T,S,A,[B]	   \\ 
0035-02  & 0.220  & BLRG & FRII  &   3   & 1.54  &   125  &  1.72 & 0.64&   109$\pm$174  & 1.62  & 118   &  B,F,[S]	   \\  
2314+03  & 0.220  & NLRG & FRII  &   1   & 1.05  &   -4   &  1.73 & 0.64&   -62$\pm$126  & 1.61  & -52   &  2F,[T]	   \\  
1932-46  & 0.231  & BLRG & FRII  &   8   & 0.52  &   645  &  1.78 & 0.66&   537$\pm$262  & 1.72  & 542   &  2F,A,I	   \\  
1151-34  & 0.258  & QSO  & CSS   &   2   & 2.38  &   -34  &  1.94 & 0.73&     5$\pm$167  & 1.82  & 16	 &  F,[S]	   \\  
0859-25  & 0.305  & NLRG & FRII  &   2   & 1.43  &   54   &  2.19 & 0.76&   -18$\pm$173  & 2.07  & -7	 &  2N  	   \\  
2250-41  & 0.310  & NLRG & FRII  &   2   & 1.80  &    19  &  2.64 & 0.77&   -61$\pm$187  & 2.77  & -74   &  2B,[T],[F]     \\ 
1355-41  & 0.313  & QSO  & FRII  &   6   & 2.53  &   332  &  2.21 & 0.76&   363$\pm$262  & 2.05  & 377   &  S,T 	   \\  
0023-26  & 0.322  & NLRG & CSS   &   9   & 1.53  &   722  &  2.26 & 0.79&   651$\pm$314  & 2.13  & 664   &  A,[D]	   \\  
0347+05  & 0.339  & WLRG & FRII  &   11  & 0.95  &   992  &  2.35 & 0.81&   853$\pm$351  & 2.28  & 860   &  B,3T,D	   \\  
0039-44  & 0.346  & NLRG & FRII  &   3   & 1.34  &   165  &  2.39 & 0.79&    61$\pm$215  & 2.31  & 69	 & 2N,3S,[T],[D]   \\  
0105-16  & 0.400  & NLRG & FRII  &   9   & 3.45  &   588  &  2.62 & 0.72&   676$\pm$348  & 2.54  & 684   &  B		   \\  
1938-15  & 0.452  & BLRG & FRII  &   5   & 2.95  &   231  &  3.16 & 1.16&   207$\pm$301  & 2.98  & 227   &  F		   \\ 
1602+01  & 0.462  & BLRG & FRII  &   5   & 2.75  &   256  &  3.18 & 1.20&   207$\pm$306  & 2.98  & 229   &  F,S,[J]	   \\ 
1306-09  & 0.467  & NLRG & CSS   &   12  & 5.08  &   792  &  3.18 & 1.22&  1008$\pm$431  & 2.97  & 1033  &  2N,S	   \\ 
1547-79  & 0.483  & BLRG & FRII  &   6   & 8.85  &   -331 &  3.23 & 1.29&   322$\pm$333  & 3.03  & 345   &  2N,T	   \\ 
1136-13  & 0.556  & QSO  & FRII  &   2   & 3.04  &   -131 &  2.74 & 0.90&   -93$\pm$247  & 2.87  & -110  &  T,J 	   \\ 
0117-15  & 0.565  & NLRG & FRII  &   9   & 5.84  &   402  &  2.72 & 0.90&   799$\pm$420  & 2.86  & 781   &  3N,S,I,[D]     \\ 
0252-71  & 0.563  & NLRG & CSS   &   6   & 2.89  &   395  &  2.72 & 0.91&   416$\pm$356  & 2.87  & 398   &  [A] 	   \\ 
0235-19  & 0.620  & BLRG & FRII  &   5   & 4.71  &    39  &  2.73 & 0.98&   306$\pm$354  & 2.79  & 298   &  2T,[B]	   \\ 
2135-20  & 0.636  & BLRG & CSS   &   7   & 2.81  &   574  &  2.71 & 1.01&   588$\pm$408  & 2.75  & 583   &  F		   \\ 
0409-75  & 0.693  & NLRG & FRII  &   11  & 1.63  &   1360 &  2.66 & 1.04&  1210$\pm$520  & 2.66  & 1210  &  2N  	   \\ 
\hline		     		  	      		 
\end{tabular}						 
\caption{Individual spatial clustering amplitudes of the 2Jy PRGs. Errors are reported for B$_{gq}^{av}$ values only, for the sake of simplicity. 
Columns 3 and 4 list the optical classification and radio morphology of the galaxies from \citet{Dicken08}.
$\sigma$ corresponds to the standard deviation of the number of background 
galaxies calculated using all the dedicated offset fields in a given filter (N$_b^{av}$ and N$_b^{med}$). Last column corresponds to the
morphological classification in \citealt{Ramos11} and \citet{Bessiere12}: T: Tail; F: Fan; B: Bridge; S: Shell; D: Dust feature;  
2N: Double Nucleus; 3N: Triple Nucleus; A: Amorphous Halo; I: Irregular feature; and J: Jet. Brackets indicate uncertain
identification of the feature.}
\label{Bgq_individual}
\end{table*}						 

\begin{table*}
\centering
\begin{tabular}{lccccccccccl}
\hline
\hline
ID          &   z     & N$_t$ & N$_b$ &  B$_{gq}$ &  N$_b^{av}$   & $\sigma$  &   B$_{gq}^{av}\pm\Delta B_{gq}^{av}$ & N$_b^{med}$ &  B$_{gq}^{med}$ & Morphology \\
 (1)        &  (2)    & (3)   & (4)   & (5)       & (6)           &   (7)     &  (8)		&  (9)        & (10)		& (11)      \\  
\hline
J0025-10    & 0.303   &  2  & \dots & \dots &  2.17 & 0.75 &   -16$\pm$175  & 2.11 & -10  &  2N,2T       \\
J0332-00    & 0.310   &  3  &  2.63 &  36   &  3.20 & 0.95 &   -19$\pm$222  & 3.11 & -11  &  2N,S,F,[B]  \\
J0234-07    & 0.310   &  1  &  1.80 &  -76  &  2.21 & 0.77 &  -115$\pm$152  & 2.09 & -104 &  \dots       \\
J0159+14    & 0.319   &  4  & \dots & \dots &  2.24 & 0.78 &   169$\pm$227  & 2.11 & 182  &  [B]         \\
J0948+00    & 0.324   &  1  &  1.77 &  -75  &  2.27 & 0.79 &  -123$\pm$155  & 2.13 & -110 &  \dots       \\
J0217-00    & 0.344   &  3  &  1.57 &  142  &  2.36 & 0.80 & 	63$\pm$212  & 2.29 & 71	&  T,I,F       \\
J0848+01    & 0.350   &  4  &  3.00 &  100  &  2.41 & 0.80 &   159$\pm$237  & 2.30 & 170  &  2S	       \\
J0904-00    & 0.353   &  7  &  3.65 &  336  &  2.38 & 0.75 &   463$\pm$295  & 2.31 & 470  &  T,S         \\
J0227+01    & 0.363   &  4  &  2.13 &  190  &  2.43 & 0.74 &   159$\pm$241  & 2.28 & 174  &  A,2S,T      \\
J0218-00    & 0.372   &  2  &  2.59 &  -61  &  2.47 & 0.74 &   -49$\pm$199  & 2.37 & -37  &  I,A,[B]     \\
J0217-01    & 0.375   &  1  &  1.89 &  -91  &  2.49 & 0.74 &  -153$\pm$170  & 2.42 & -146 &  \dots       \\
J0924+01    & 0.380   &  2  &  1.92 &	8   &  2.53 & 0.73 &   -55$\pm$200  & 2.44 & -45  &  T,[B]       \\
J0320+00    & 0.384   &  15 &  2.53 &  1296 &  2.53 & 0.73 &  1297$\pm$425  & 2.46 & 1304 &  I,[S]       \\
J0923+01    & 0.386   &  5  &  2.06 &  307  &  2.55 & 0.73 &   256$\pm$271  & 2.49 & 261  &  S,F,[T]     \\
J0142+14    & 0.389   &  4  & \dots & \dots &  2.57 & 0.73 &   149$\pm$251  & 2.49 & 158  &  \dots       \\
J0114+00    & 0.389   &  5  &  2.56 &  255  &  2.57 & 0.73 &   254$\pm$272  & 2.49 & 262  &  2N,S        \\
J0123+00    & 0.399   &  9  &  1.85 &  756  &  2.61 & 0.71 &   676$\pm$348  & 2.59 & 679  &  2N,B,[A]    \\
J2358-00    & 0.402   &  3  &  2.70 &  32   &  2.62 & 0.73 & 	40$\pm$231  & 2.60 & 43	&  B,T,F       \\
J0334+00    & 0.407   &  3  &  2.11 &  95   &  2.66 & 0.73 & 	36$\pm$235  & 2.66 & 37	&  S	       \\
J0249+00    & 0.408   &  1  &  2.11 & -118  &  2.65 & 0.73 &  -177$\pm$180  & 2.65 & -176 &  S,B         \\
\hline		     			      		 
\end{tabular}						 
\caption{Same as in Table \ref{Bgq_individual} but for the type-2 quasars studied in \citet{Bessiere12}.}
\label{Bgq_individual2}
\end{table*}

\section{Results}

Here we present the results of the study of the environments of luminous radio-loud and radio-quiet quasars. 
In Table \ref{Bgq} we report mean values of 
B$_{gq}$, B$_{gq}^{av}$ and B$_{gq}^{med}$ and standard errors ($\sigma(B_{gq})/\sqrt{n}$, with n equal to 
the number of targets included in the mean) for these groups. 
As explained in Section \ref{cluster}, we consider B$_{gq}^{av}$ 
more reliable than B$_{gq}$ and B$_{gq}^{med}$ 
because we have measurements of N$_b^{av}$ for all the PRGs and type-2 quasars (see Tables \ref{Bgq_individual}
and \ref{Bgq_individual2}). 
Therefore, the results discussed below were obtained using B$_{gq}^{av}$ unless otherwise stated. For the sake of 
simplicity, we only report individual errors for the B$_{gq}^{av}$ values in Tables \ref{Bgq_individual}, 
\ref{Bgq_individual2}, \ref{Bgq_control_individual} and \ref{Bgq_control_individual2}.

\begin{figure*}
\centering
\subfigure[]{\includegraphics[width=8.7cm]{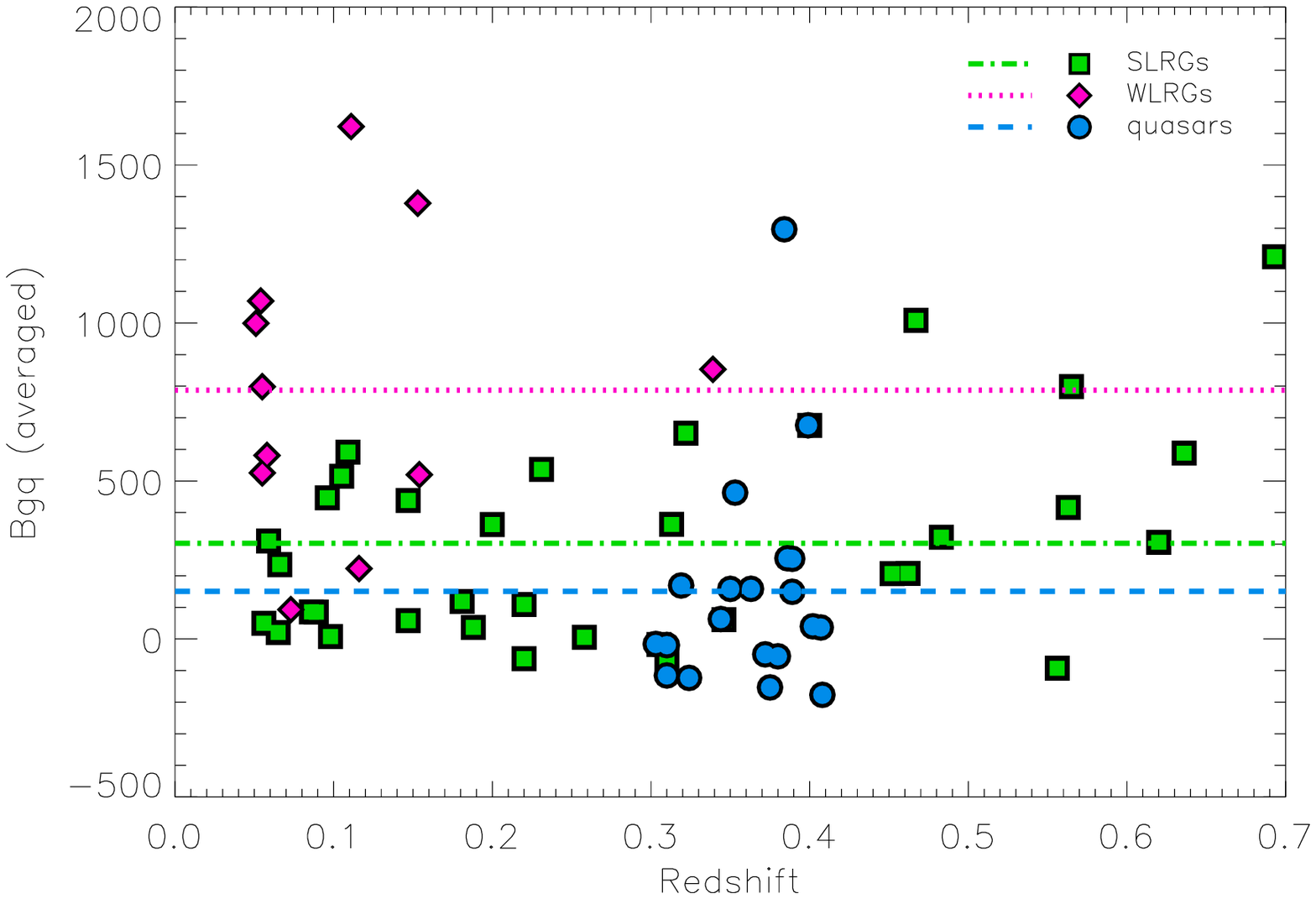}
\label{redshift}}
\subfigure[]{\includegraphics[width=8.7cm]{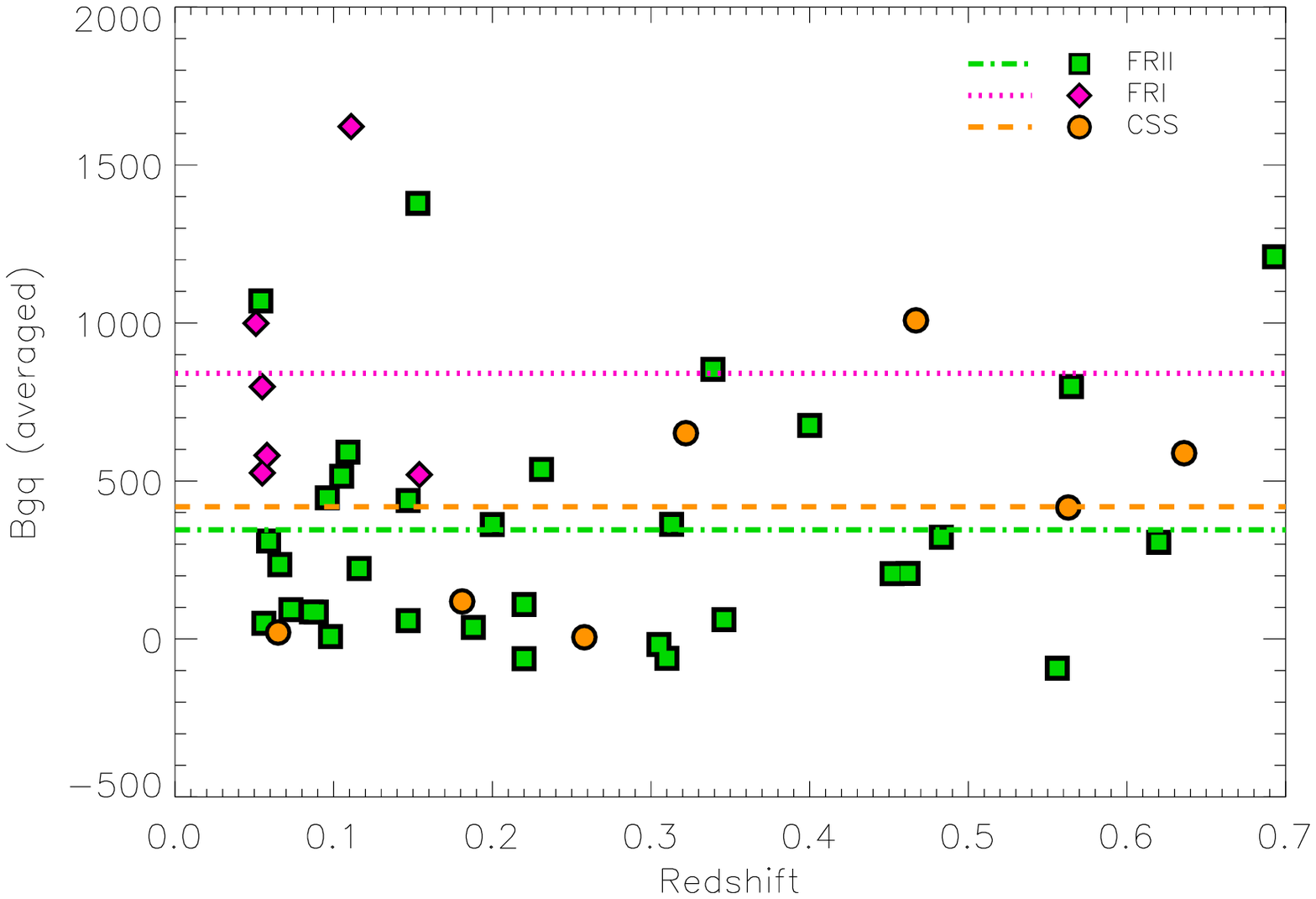}
\label{redshift_FR}}
\subfigure[]{\includegraphics[width=8.7cm]{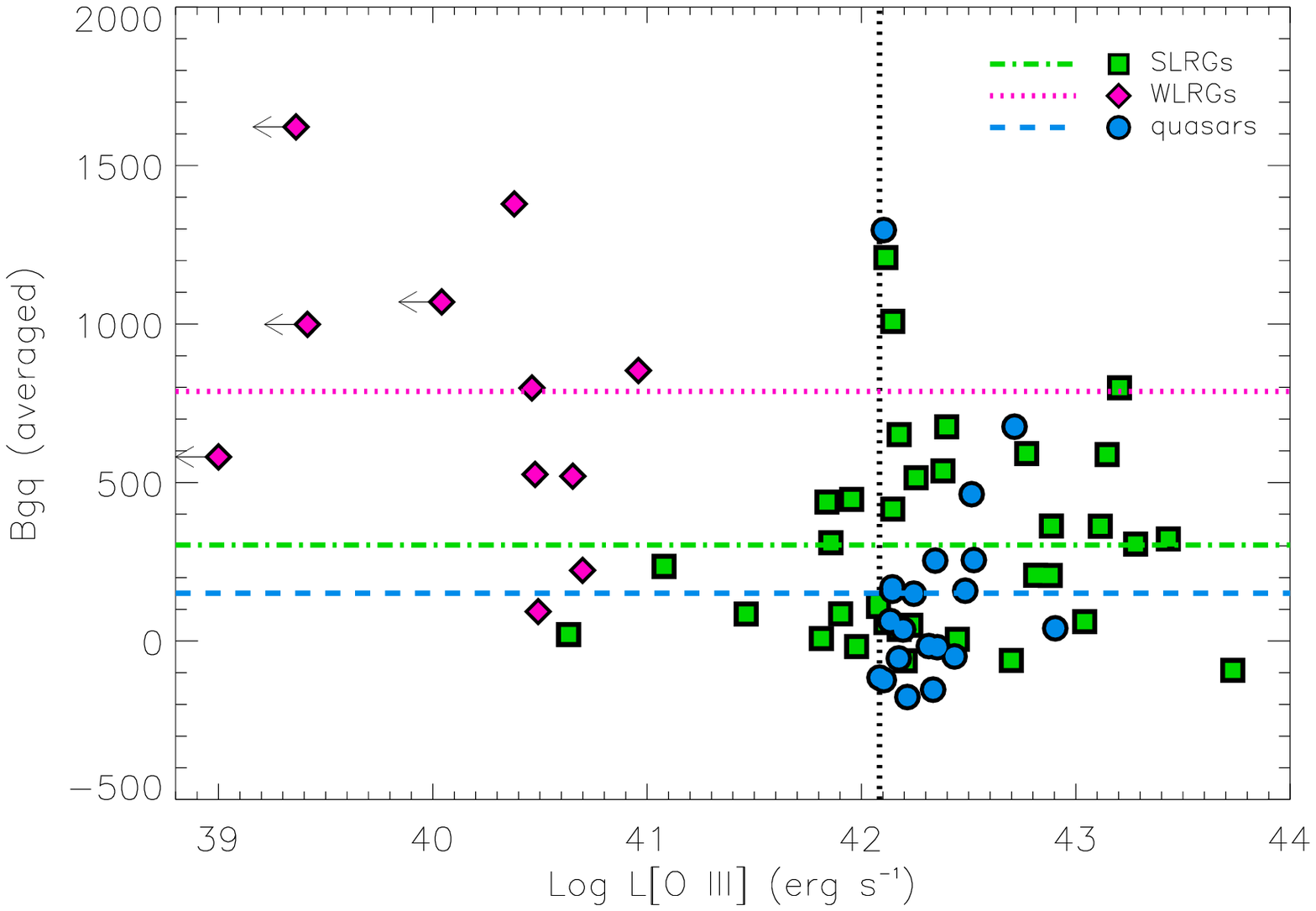}
\label{lO3}}
\subfigure[]{\includegraphics[width=8.7cm]{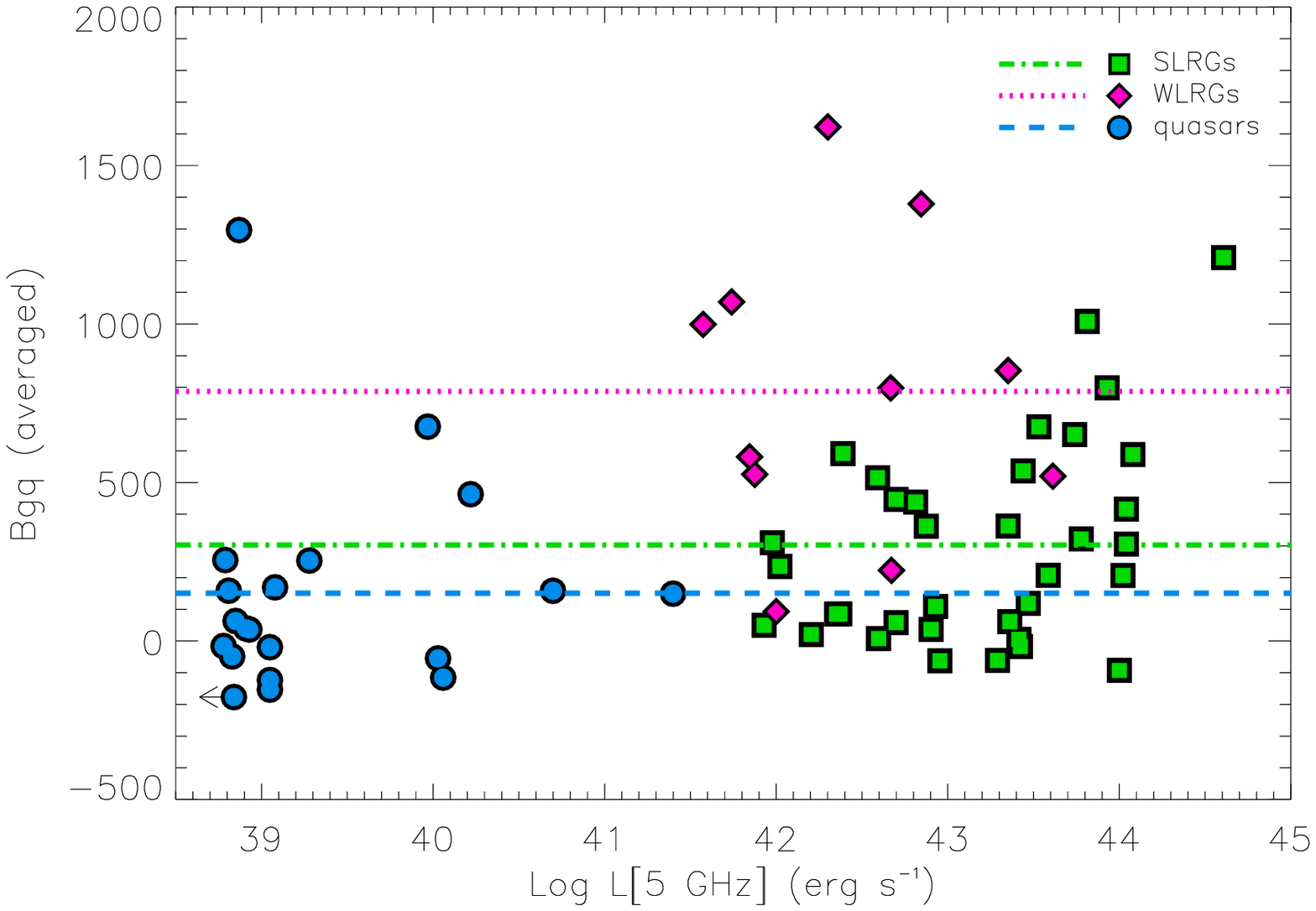}
\label{l5GHz}}
\caption{(a) Spatial clustering amplitude versus redshift for the SLRGs, WLRGs and type-2 quasars in the samples considered here. 
Horizontal lines correspond to the mean B$_{gq}^{av}$ of each class (dashed line: type-2 quasars, dotted line: 
WLRGs and dot-dashed line: SLRGs). 
(b) Same as in Figure \ref{redshift}, but for PRGs only and using their radio classification: 
FRIs (dotted line), FRIIs (dot-dashed line) and CSS/GPS (dashed line).
(c) Same as in Figure \ref{redshift}, but versus [O III]$\lambda$5007 luminosity (in units of $\nu L_{\nu}$).
Dotted vertical line indicates Log (10$^{8.5} L_{\sun}$).
(d) Same as in Figure \ref{lO3} but versus 5 GHz luminosity (in units of $\nu L_{\nu}$).
\label{z}}
\end{figure*}

Figure \ref{z} summarises the individual B$_{gq}^{av}$ results, where they are plotted against redshift, [O III]$\lambda$5007 
emission line luminosity and radio power for the different groups considered in this work. 
The [O III]$\lambda$5007 integrated luminosities and 5GHz monochromatic luminosities were taken
from Table 1 in \citet{Dicken09}, and transformed into $\nu L_{\nu}$ luminosities.



\subsection{Abell classification}
\label{abell}

To better compare the results of our study of the environment of radio-loud and radio-quiet AGN with the 
literature, here we provide an estimation of the typical spatial clustering amplitudes for 
the five Abell richness classes, in the chosen cosmology. 

As described by \citet{McLure01}, the correlation between B$_{gq}$ and Abell class is affected by a large scatter, and
thus, there is no rigorous transformation between them. Here we have adopted the linear squeme employed by
\citet{Yee99} and \citet{McLure01}, in which the different Abell classes are separated by $\Delta B_{gq}$ = 400 Mpc$^{1.77}$. 
We use the same normalization as in \citet{McLure01}, that re-calibrated to our cosmology corresponds 
to $B_{gq}\sim$400 Mpc$^{1.77}$. Therefore, for Abell classes 0, 1, 2, 3, 4 and 5, we have B$_{gq}$ = 400, 800, 1200, 
1600, 2000 and 2400 respectively.

To check the transformation between B$_{gq}$ and Abell class, we can look at the 2Jy PRGs that are known to be at the centre 
of galaxy clusters, and see if they have B$_{gq}$ values $\geq$400 (i.e., Abell class 0 or higher).
There are at least four PRGs in clusters, according to the literature:

\begin{itemize}

\item PKS 0620-52. A cluster environment for this radio galaxy is supported by the existence 
of a moderately luminous X-ray halo, for which \citet{Trussoni99} estimated a 0.5-3.5 keV luminosity of 
2.0$\times10^{44}$ erg~s$^{-1}$, once transformed to our chosen cosmology. 

\item PKS 0625-35 was the first ranked member of the cluster A3392 \citep{Trussoni99}. \citet{Siebert96} measured
a X-ray luminosity of 2.3$\times10^{44}$ erg~s$^{-1}$ in the 0.1-2.4 keV band for the extended halo of this source. 

\item PKS 0915-11 (Hydra A) is situated in the Hydra cluster of galaxies and it is one the most powerful radio sources in the 
local universe. \citet{McNamara00} reported the discovery of structure in the central 80 kpc of the cluster X-ray-emitting gas, 
of 0.5-4.5 keV luminosity 2.2$\times10^{44}$ erg~s$^{-1}$.
More recently, \citet{Wise07} claimed the existence of an extensive cavity system, as revealed from a deep Chandra image of 
the hot plasma. 

\item PKS 1648+05 (Herc A) is at the centre of a cooling flow cluster of galaxies at z=0.154. The X-ray luminosity 
of the cluster 2.7$\times10^{44}$ erg~s$^{-1}$ in the 0.1-2.4 keV band \citep{Siebert99}. 
A recent analysis of Chandra X-ray data showed that the cluster has cavities and a shock front associated with the radio source 
\citep{Nulsen05}.

\end{itemize}

These four galaxies have spatial clustering amplitudes (B$_{gq}^{av}$) of 999, 526, 798 and 520 respectively, 
which, according to our calibration, correspond to Abell classes 1 and 0.  Thus, in 
the following, we can consider values of B$_{gq}^{av}\ga 400$ typical of cluster environments. 

\subsection{WLRGs versus SLRGs}

In Figure \ref{redshift} we plotted the spatial 
clustering amplitude (B$_{gq}^{av}$) versus redshift for the SLRGs (green squares), the WLRGs (pink diamonds) and 
the type-2 quasars (blue circles). In general, WLRGs are concentrated at lower redshifts
and are in denser enviroments than SLRGs and type-2 quasars. 

\begin{figure*}
\centering
{\par
\includegraphics[width=6.7cm]{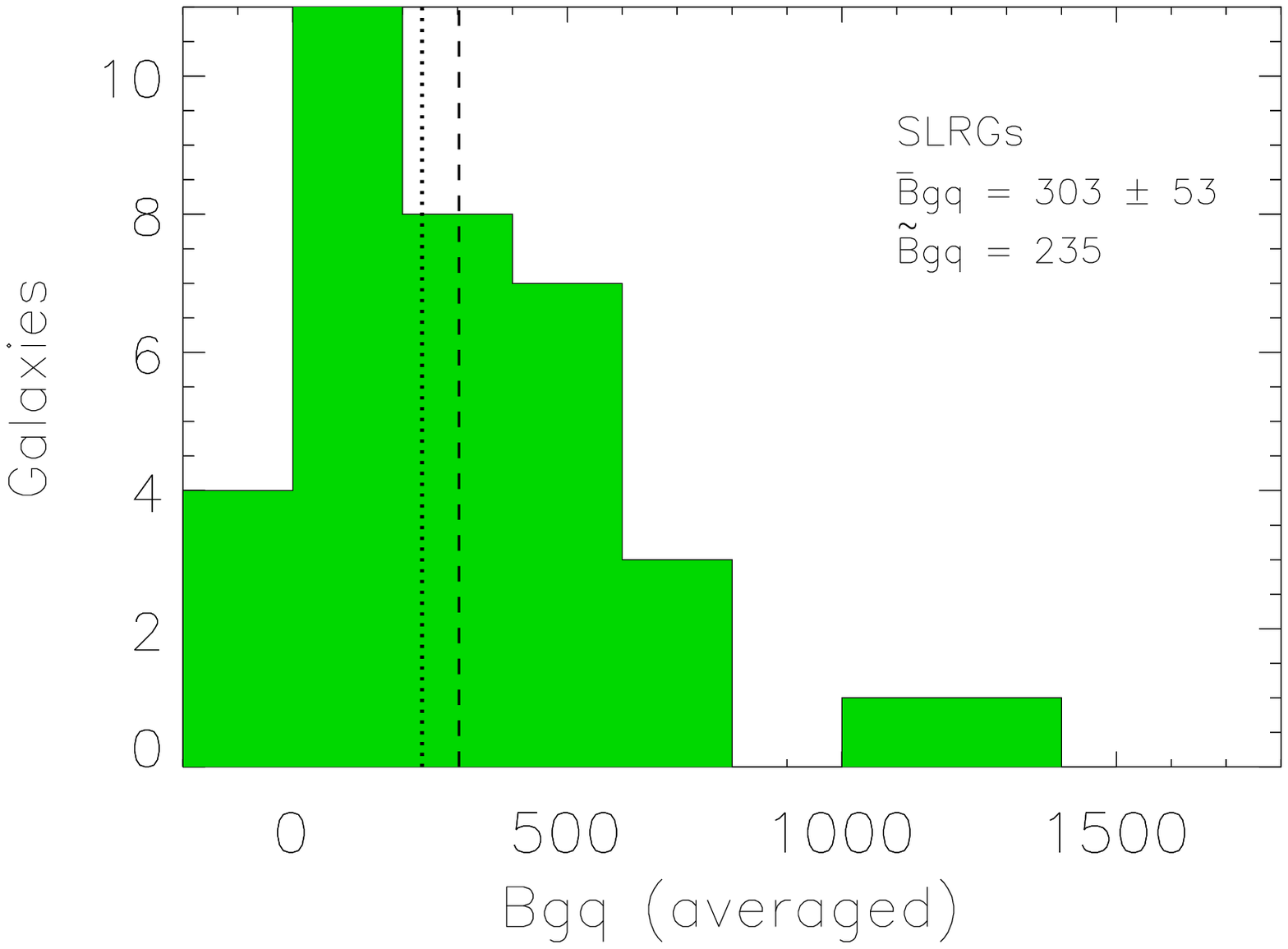}
\includegraphics[width=6.7cm]{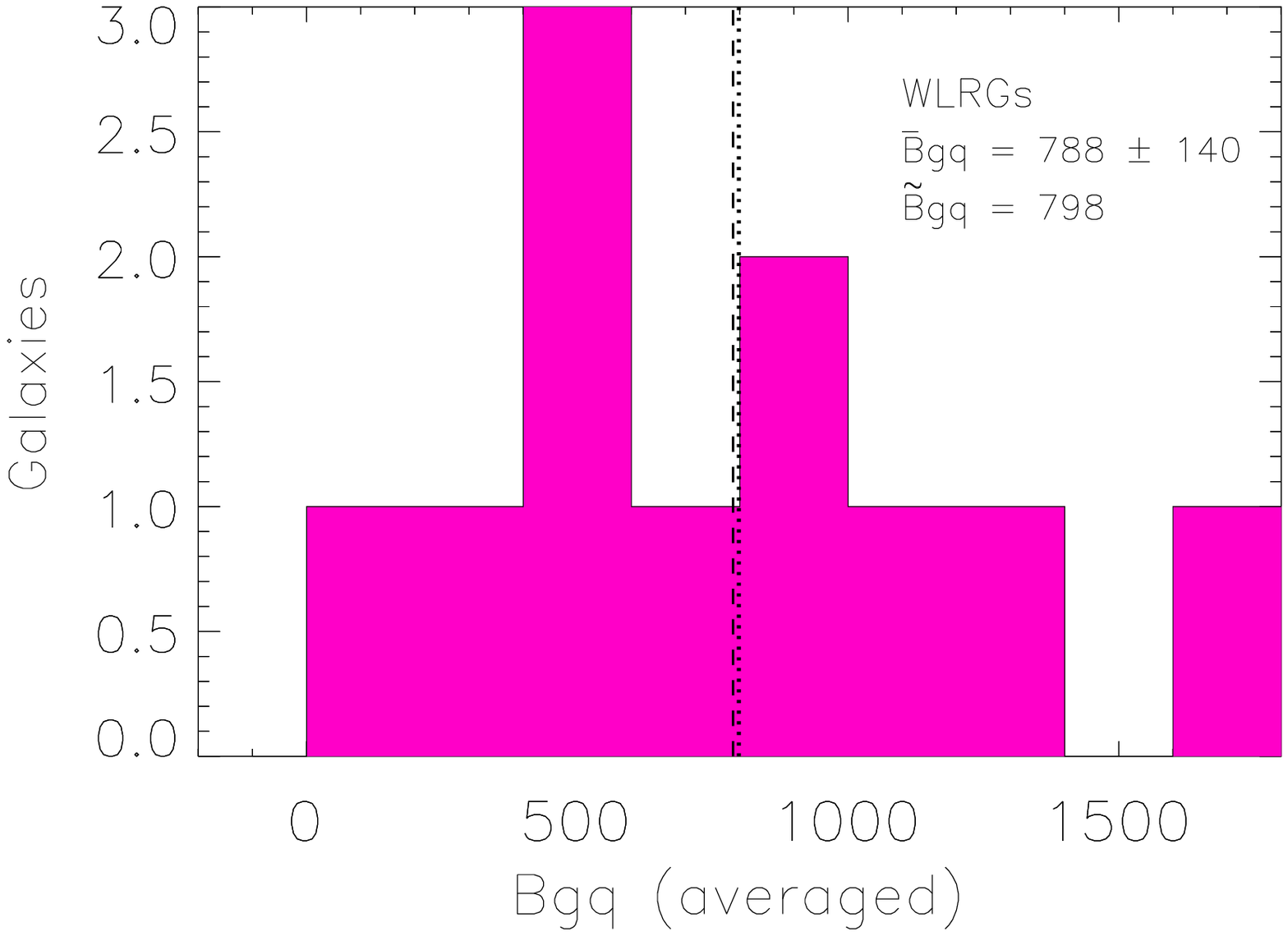}
\includegraphics[width=6.7cm]{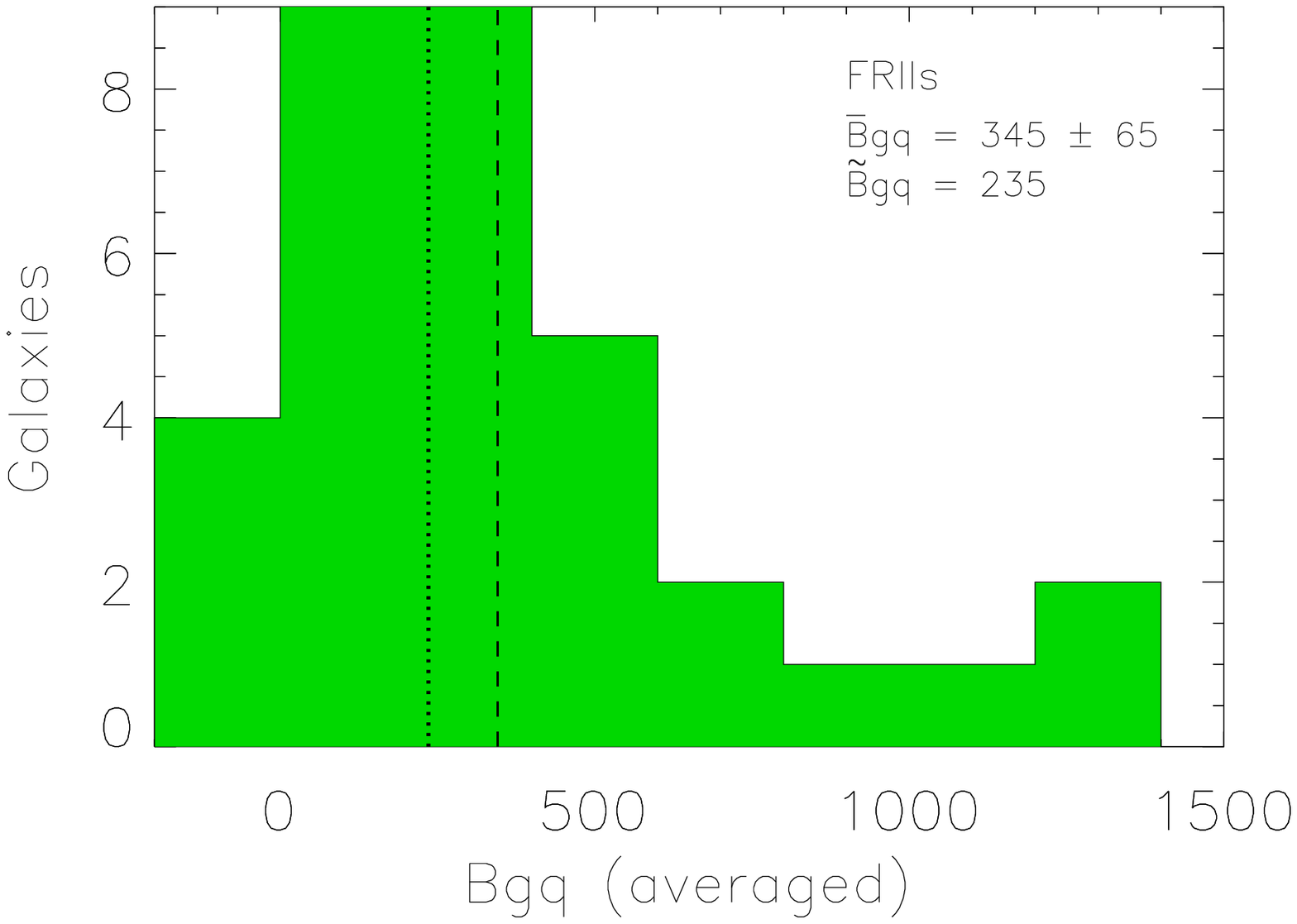}
\includegraphics[width=6.7cm]{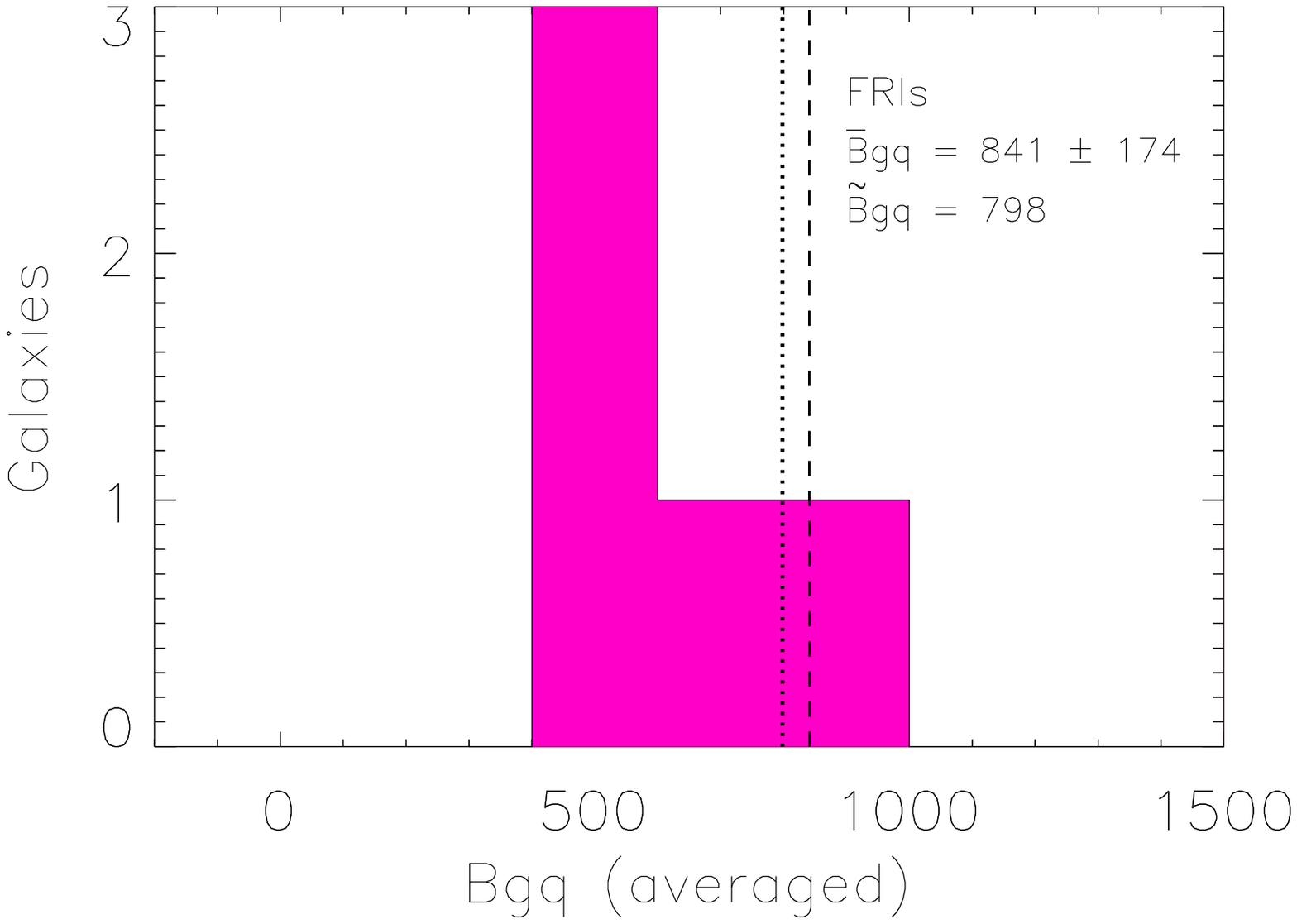}
\includegraphics[width=6.7cm]{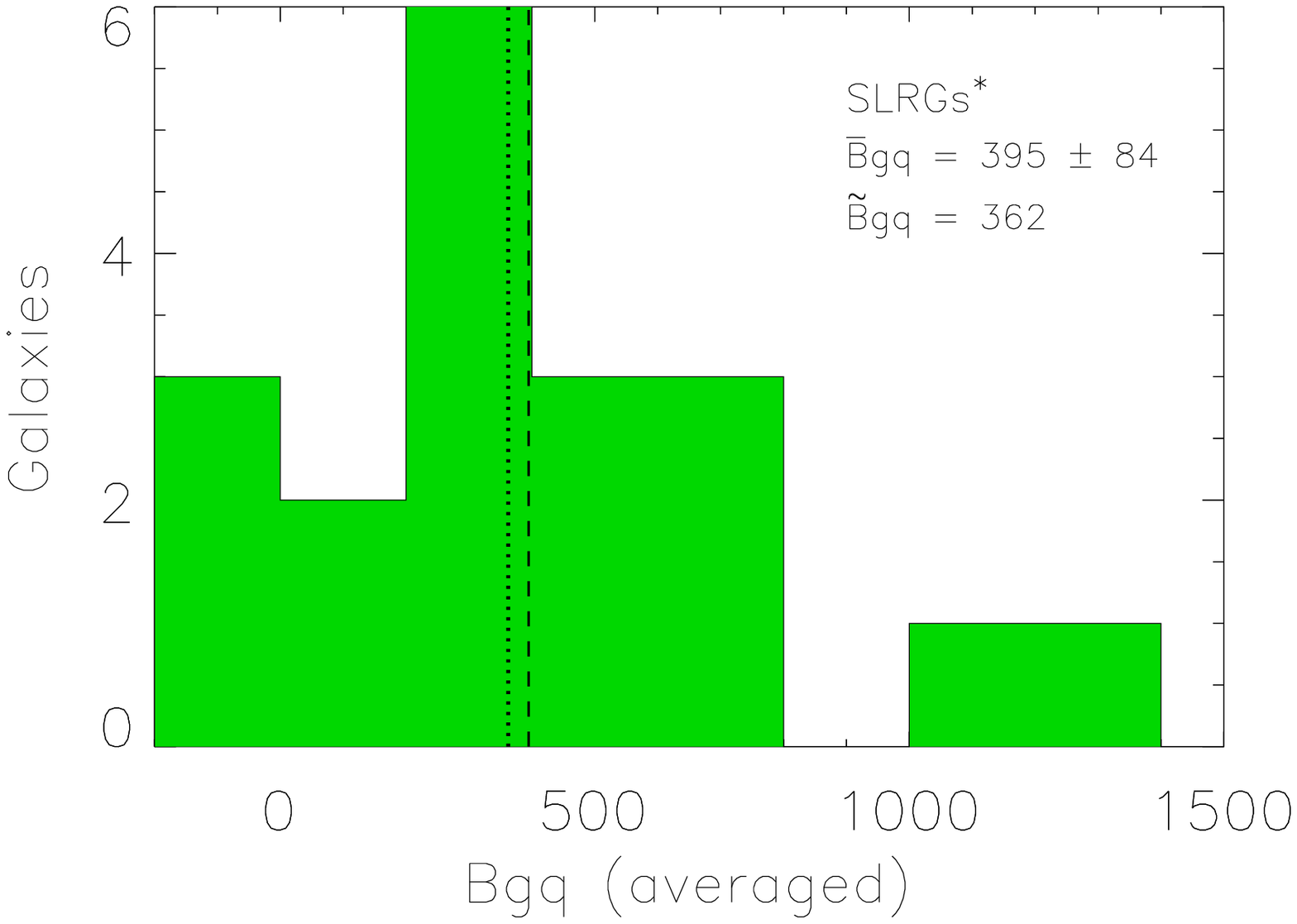}
\includegraphics[width=6.7cm]{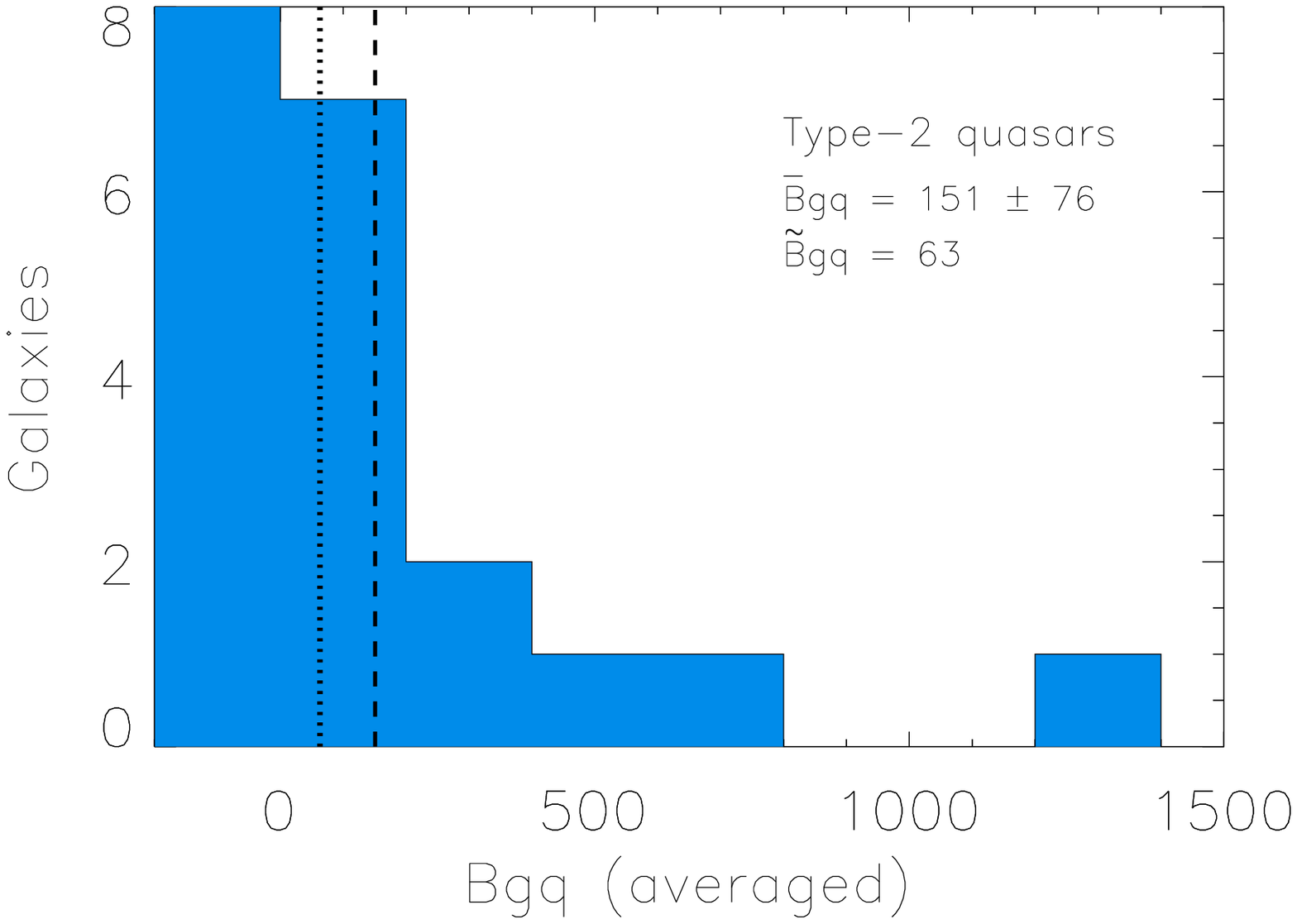}\par}
\caption{Distribution of spatial clustering amplitudes for the SLRGs and WLRGs (top panels),
FRII and FRI radio galaxies (middle panels) and SLRGs* and type-2 quasars (bottom panels). 
Dotted and dashed lines correspond to the median ($\tilde{B}_{gq}^{av}$) and the average ($\bar{B}_{gq}^{av}$) 
of the distribution, which are also reported at the top of each panel. Colours are the same as in Figure \ref{z}.
\label{hist1}}
\end{figure*}

We used the Kolmogorov-Smirnov (KS) test to compare the distributions of B$_{gq}^{av}$ of the 2Jy WLRGs 
and SLRGs, shown in the top panels of Figure \ref{hist1}. 
We found that WLRGs are in richer environments than SLRGs, with mean clustering amplitudes of 
$\bar{B}_{gq}^{av}(SLRGs)=303\pm53$ and $\bar{B}_{gq}^{av}(WLRGs)=788\pm140$, and this
difference is significant at the 3$\sigma$ level (see Table \ref{Bgq}). 

Since the redshift distributions of WLRGs and SLRGs are quite different, we compared the environments
of the two groups only considering galaxies at $z<0.2$. By doing this redshift cut, we have 
14 SLRGs with mean clustering amplitude $\bar{B}_{gq}^{av}=214\pm55$ and 10 WLRGs with $\bar{B}_{gq}^{av}=781\pm155$.
As in the case of the comparison done considering the whole redshift range, this difference is significant at the 3$\sigma$ level,
based on the KS test.

Of the 11 WLRGs, all but PKS 0034-01 and PKS 0043-42\footnote{Note that recently, 
based on mid-infrared Spitzer spectroscopic data, \citet{Ramos11b} claimed that PKS 0043-42 has a dusty torus, 
which is a feature typical of SLRGs.} have individual B$_{gq}^{av}$ values characteristic of Abell 0, 1, 2 and 3 clusters,
which are larger than the mean value of the whole PRG sample ($\bar{B}_{gq}^{av}=419\pm60$). According to this,
WLRGs constitute a different class of PRGs on the basis of both their spatial clustering amplitude and their
optical classification \citep{Tadhunter98}.

The case of the SLRGs is different. There are only 12 SLRGs with B$_{gq}^{av}\ga 400$, of which nine have
clustering amplitudes characteristic of Abell class 0. The other three SLRGs are PKS 1306-09 and 
PKS 0117-15 (Abell class 1) and PKS 0409-75 (Abell class 2).
Summarising, 82\% of the WLRGs in the 2Jy sample are in clusters, according to our definition (B$_{gq}\ga 400$),
compared with only 31\% of the SLRGs.

The lack of disturbed morphologies in 73\% of the 2Jy WLRGs (\citealt{Ramos11}), and their large clustering amplitudes, 
may indicate that at least some WLRGs could be powered by a different triggering mechanism, either cooling flows 
sinking towards the cluster centers \citep{Tadhunter89,Baum92,Bremer97,Edge99,Edge10,Best05,Sabater13} or direct 
accretion of hot gas from the X-ray haloes \citep{Best06,Hardcastle07}.
However, we must be cautious about possible 
observational selection effects. In particular, it is more difficult to detect 
tidal features such as shells or broad fans in regions 
of high galaxy density, since the tidal effects rapidly disrupt these features (see \citealt{Linden10} and
references therein). 
Regarding the 27\% of 2Jy WLRGs showing the tidal features, 
it is well below the background rate of interactions measured for the quiescent population of early-type 
galaxies of same mass and redshift (53\%; \citealt{Ramos12}). Thus, the galaxy interactions occurring 
in those WLRGs may or may not be linked to the fuelling of the AGN.

\subsection{FRIs versus FRIIs}

In Figure \ref{redshift_FR} we show the individual B$_{gq}^{av}$ values of the 2Jy PRGs plotted against redshift 
highlighting their classification at radio wavelengths. Green squares correspond to FRIIs (33 objects), pink diamonds to 
FRIs (6 objects), and orange circles to CSS/GPS sources (7 objects). 

The FRIs in the sample have redshifts $z<0.2$ and the majority have larger
values of B$_{gq}$ than FRIIs and CSS/GPS sources. In fact, their mean clustering amplitude (B$_{gq}^{av}=841\pm174$) is 
characteristic of an Abell class 1 cluster. This result is not surprising, considering that 
all the FRIs in the 2Jy sample are WLRGs. Interestingly, WLRGs (and consequently, FRIs) tend to have large 
ratios of radio luminosity to AGN power, constituting a first indication that dense environments may 
boost the radio emission of PRGs (see Section \ref{radiopower} for further discussion on this).

In the two central panels of Figure \ref{hist1}, we compare the spatial clustering amplitudes of the
FRI and FRII radio galaxies in the 2Jy sample. The distributions, based on the KS test, are different at the 3$\sigma$ 
level if we consider B$_{gq}^{av}$ (see Table \ref{Bgq}).
This is in agreement with the results found by \citet{Gendre13}, based on a sample of $\sim$200 radio galaxies
at redshift $z\leq$0.3 (see also \citealt{Prestage88,Prestage89,Zirbel97}). 

\citet{Hill91} studied the cluster environments of a sample of 45 FRII radio galaxies at z$\sim$0.5\footnote{Including 
members of the 3CRR, 1Jy, 5C12 and LBDS samples. See \citet{Hill91} and references therein.} and compared 
them with their low-redshift counterparts. Based on this comparison, \citet{Hill91} claimed that high-redshift PRGs are in 
richer environments than those at low-redshift. 
However, looking at Figure \ref{redshift_FR}, we do not observe an 
enhancement in the clustering amplitude of FRIIs with redshift. In fact, if we divide the FRIIs into a low-redhift 
sample ($z<0.2$; 16 sources) and high-redshift sample ($0.2\leq z<0.7$; 17 sources), we do not find 
a significant trend in the environments with redshift: $\bar{B}_{gq}^{av}$($z<0.2$) = 351$\pm$98 and 
$\bar{B}_{gq}^{av}$($0.2\leq z<0.7$) = 340$\pm$90.
A lack of redshift dependence in B$_{gq}$ was also reported by \citet{Wold00}, based on the comparison of 
a sample of 21 radio-loud quasars with redshifts $0.5\leq z\leq0.82$ with other literature samples at lower redshifts.
Also \citet{McLure01} reported no epoch dependence in the environments of radio-loud and radio-quiet powerful AGN out to 
redshift z=0.5. 



\begin{table}
\centering
\footnotesize
\begin{tabular}{lcccc}
\hline
\hline
Comparison   &  Targets & $\bar{B}_{gq}$    & $\bar{B}_{gq}^{av}$   &  $\bar{B}_{gq}^{med}$ \\
\hline
SLRGs	     &  35      & 233$\pm$64  & 303$\pm$53      &  319$\pm$58	\\ 	
WLRGs 	     &  11      & 759$\pm$156 & 788$\pm$140     &  795$\pm$253	\\ 
KS test      &  \dots   & 98.2\%      & 99.7\%          &  78.6\% 	\\ 
\hline
FRII         &   33     & 284$\pm$75  & 345$\pm$65      &  344$\pm$69	\\ 
FRI          &   6      & 811$\pm$230 & 841$\pm$174     & 1087$\pm$551  \\
KS test      &  \dots   & 98.8\%      & 99.8\%          &  \dots        \\
\hline		     			      		 
SLRGs*       &  19      & 337$\pm$90  & 395$\pm$84      &  400$\pm$84	\\ 	
Type-2 quasars& 20      & 184$\pm$87  & 151$\pm$76      &  159$\pm$76	\\ 
KS test      &  \dots   & 83.9\%      & 98.8\%          &  98.8\% 	\\ 
\hline
\end{tabular}						 
\caption{Comparison between the cluster environments of 1) SLRGs and WLRGs, 2) FRIIs and FRIs, and 3) SLRGs and type-2 quasars with 
[O III] luminosities larger than 10$^{8.5}L_{\sun}$.
$\bar{B}_{gq}\pm\sigma(B_{gq})/\sqrt{n}$ is reported for each group, with n equal to the number of targets included in the mean, 
together with the results of the KS test (significance level). We do not report KS test results for the comparison between 
$B_{gq}^{med}$(FRI) and $B_{gq}^{med}$(FRII) because there are only two FRIs with $B_{gq}^{med}$ available.}
\label{Bgq}
\end{table}


\subsection{PRGs and type-2 quasars}

The type-2 quasars are concentrated around low values of B$_{gq}$ (see Table \ref{Bgq}), with the 
exceptions of J0904-00, J0320+00 and J0123+00 (B$_{gq}\ga 400$; i.e. cluster-like). 
To compare the environments of PRGs and type-2 quasars, it 
is necessary to consider the same selection criterion used by \citet{Bessiere12} for the type-2 quasars, 
and select only PRGs with [O III] luminosities larger than 10$^{8.5}L_{\sun}$.
We did not only consider PRGs with redshits in the same range as the type-2 quasar sample 
(i.e. $0.3\leq z\leq 0.41$) because that would leave us with five PRGs only, 
not enough for any statistical comparison. However, we used our 
$0.2\leq z<0.7$ PRG sample to have a more comparable redshift range. 
By applying these luminosity and redshift cuts, we ended up with 19 SLRGs (hereafter SLRGs*) whose environments 
are denser, on average, than those of the 20 type-2 quasars (see bottom panels of Figure \ref{hist1}). 
The significance of this difference is 98.8\% according to the KS test (2$\sigma$; see Table \ref{Bgq}). If we further restrict the 
redshift range (e.g. $0.2\leq z<0.5$) in order to better match that of the type-2 quasars, the difference 
between environments becomes smaller (93.6\%). Thus, although the results presented here
hint at a difference between the environments of PRGs and type-2 quasars, larger samples are required to 
confirm them statistically.


If confirmed for a larger sample, the latter results would be in agreement with the pioneering works 
of \citet{Yee84,Yee87} and \citet{Ellingson91}.
More recently, using a sample of over 2,000 radio-loud AGN selected from SDSS with redshifts 0.03$<$z$<$0.3, 
\citet{Best05} claimed that optical AGN and radio-loud AGN are different phenomena and are triggered by different 
mechanisms. The latter authors claimed that the probability of a galaxy being radio-loud is independent of its classification in 
the optical. \citet{Best05} and also \citet{Kauffmann08}, using the same galaxy sample, reported that radio-loud AGN 
are generally found in denser environments than radio-quiet AGN, coinciding with our results 
(see also Inskip et al.,~in preparation for a detailed study of the host galaxy properties of the 2Jy radio galaxies). 
However, it is worth noting that the radio-loud AGN studied by \citet{Best05} have much lower radio luminosities 
(L$_{1.4GHz}=10^{23}-10^{25}$ W~Hz$^{-1}$) than the majority of 2Jy radio galaxies. 

Similar results were found at higher redshift by \citet{Donoso10} and \citet{Falder10}, based 
on samples of radio-loud and radio-quiet AGN 
at redshift 0.4$<$z$<$0.8 and z$\sim$1 respectively: both found evidence for increasing 
overdensity with increasing radio luminosity (see also \citealt{Serber06}), as well as for 
radio-loud AGN being in denser environments than radio-quiet galaxies.

On the other hand, \citet{McLure01} and \citet{Wold01} found no significant difference between the environments 
of luminous radio-loud and and radio-quiet type-1 and type-2 quasars at z$\sim$0.2. 


If confirmed, the difference between the environments of PRGs and type-2 quasars would not support the hypothesis
of luminous AGN cycling between radio-loud and radio-quiet phases within a single quasar triggering 
event (see e.g. \citealt{Nipoti05}). 
Typically, the radio-loud phase in PRGs is expected to last over a period of $t_{PRG} \sim 100$~Myr 
\citep{Leahy89,Blundell99,Shabala08},
not sufficient for a change in the large scale environment surrounding a typical radio-loud AGN. 
However, as discussed above, observations of larger samples 
are required to put these results on a firmer statistical footing.




\subsection{Star formation versus environment}

Using multi-wavelength data of the 2Jy sample of PRGs, including optical spectroscopy and mid- and 
far-infrared imaging and spectroscopy, \citet{Dicken12} searched for recent star formation activity (RSFA) in the 
host galaxies of the 46 radio sources. 
The authors used four different diagnostic methods to determine whether or not there is recent star formation 
present in the 2Jy host galaxies and they confirmed the presence of RSFA in 20\% of the sample (i.e. in nine of the 2Jy 
PRGs). Here we consider that an object has RSFA if it shows evidence for star formation activity based on a
minimum of two diagnostic methods. In \citealt{Ramos11}, we 
searched for a possible relation between optical morphology and star formation activity, but we did not
find any significant difference between the morphologies of the star-forming galaxies and those without 
recent star formation.

Now we can look at the individual spatial clustering amplitudes of the 2Jy galaxies with and without RSFA. We find that 
78\% of the galaxies with RSFA (7 of the 9) are in clusters of Abell types 0, 1 and 2. On the other hand, if we look at
the clustering amplitudes of the 35 galaxies without RSFA (we discarded the 9 PRGs with confirmed RSFA and another 2 with 
RSFA confirmed by one diagnostic method only; \citealt{Dicken12}), we find 37\% in clusters. 
Thus, in spite of the limited number of PRGs with signs of RSFA, our results show an enhancement of 
star formation activity in denser environments. 



Galaxy interactions could be an explanation 
for the detection of RSFA in the seven 2Jy PRGs in clusters. The moderate densities of these clusters 
favour galaxy interactions, and indeed we detect signs of interactions in 6 of them, 
as indicated in Table \ref{Bgq_individual}. These interactions could be leading to an enhancement 
of the star formation activity in their galaxy hosts.  
An alternative explanation for the RSFA detected in the 2Jy PRGs in relatively dense environments could 
be cooling flows taking place at the centers of these galaxy clusters.
Searching for cooling gas at the centers of galaxy clusters is very challenging because of the low 
gas density, and because these flows are much less massive than expected \citep{Fabian94,Fabian12}.
AGN feedback has been proposed as the energetic process necessary to balance radiative cooling, preventing 
massive cooling flows and intense star formation. However, very recently, \citet{McDonald12} reported the 
existence of a massive and X-ray luminous cluster at redshift z=0.6 with a cooling rate of 3820 M$_{\sun}~yr^{-1}$.
Interestingly, the central galaxy hosts a powerful AGN and a massive starburst, where stars are forming at a rate
of 740 M$_{\sun}~yr^{-1}$. \citet{McDonald12} claimed that this cluster might be an example of a system in which the AGN feedback, which
would otherwise suppress the cooling flow, is not completely established.

\section{Discussion}

In this section we discuss the differences and similarities found among the environments of our complete 
samples of PRGs, type-2 quasars and quiescent early-type galaxies (EGS and EGS*).
To perform those comparisons, we have used the KS non-parametric test for the equality of the one-dimensional distributions
of spatial clustering amplitudes. In this regard, the reader should bear in mind that, although we find significant differences 
between the environments of some of the groups discussed here, there are also substantial overlaps between them (see e.g. Figure
\ref{hist1}).

\subsection{Dependence of radio power on environment}
\label{radiopower}

\begin{figure*}
\centering
\subfigure[]{\includegraphics[width=8.5cm]{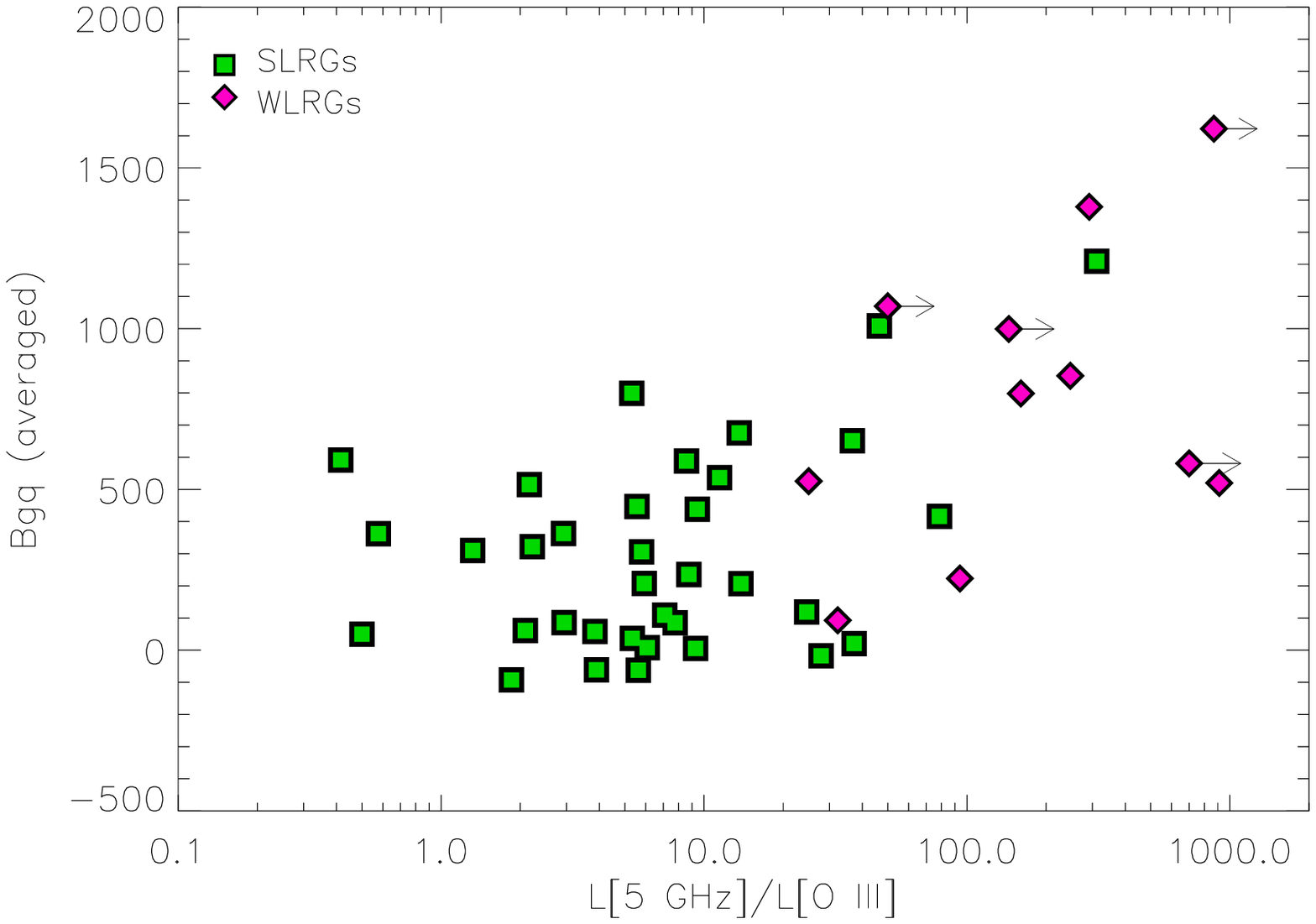}
\label{efficiency}}
\subfigure[]{\includegraphics[width=8.5cm]{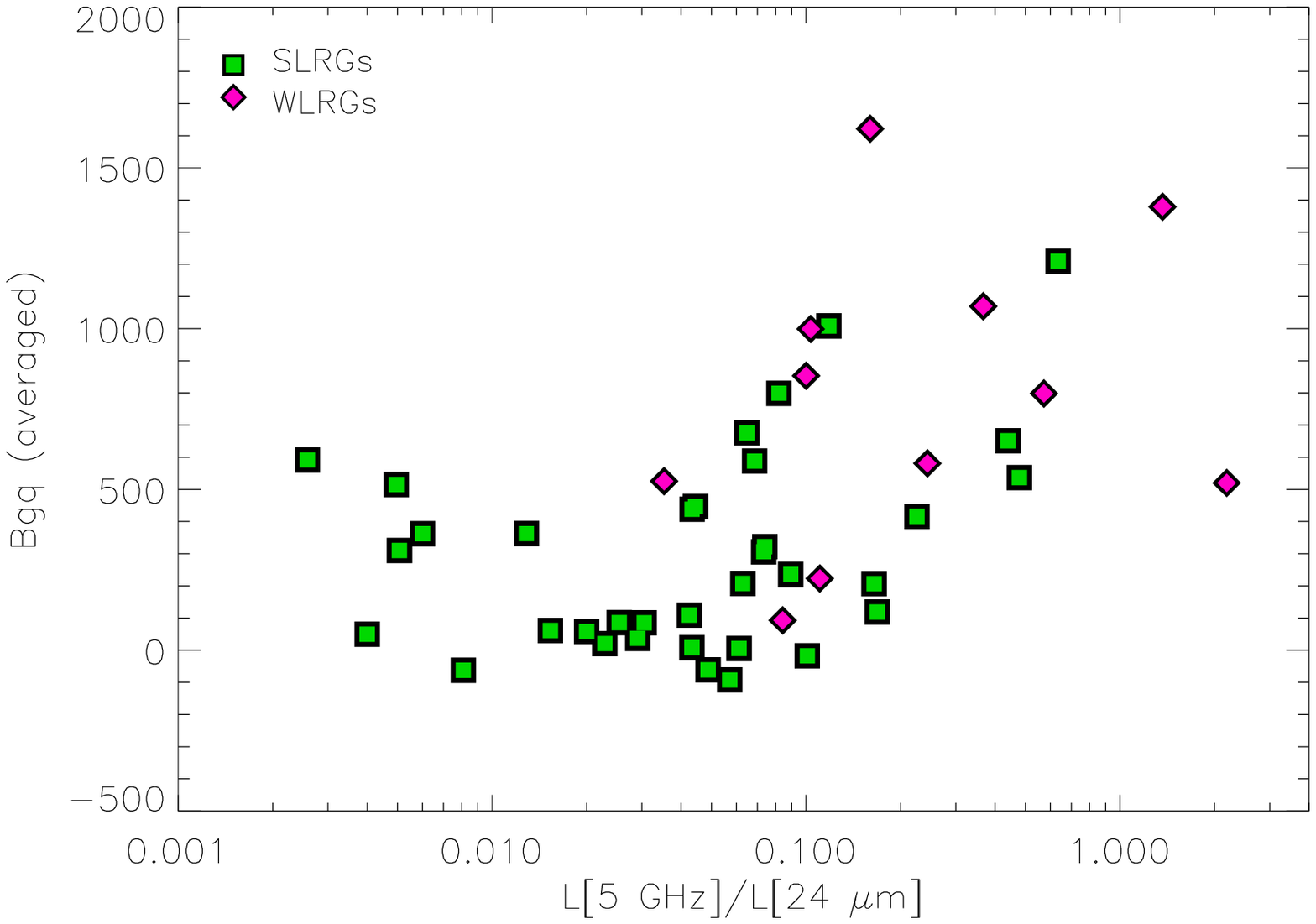}
\label{efficiency24}}
\caption{(a) Spatial clustering amplitude versus 5 GHz/[O III]$\lambda$5007 luminosity ratio for SLRGs and WLRGs.
(b) Same as in Figure \ref{efficiency} but versus 5 GHz/24 \micron. Both ratios have been calculated
using luminosities in units of $\nu L_{\nu}$. 
\label{luminosity_ratios}}
\end{figure*}

As first suggested by \citet{Barthel96} for the case of Cygnus A and a few other sources, the radio luminosity 
may be affected by the environments of the radio sources (see also \citealt{Best05,Kauffmann08,Falder10} and references therein). 
In particular, for a given intrinsic jet power, the radio luminosities 
of FRII radio galaxies may be boosted in rich environments because of the strong interaction between the 
relativistic plasma and the hot X-ray emitting gas. 
Therefore, one would expect that the richer the environment, the higher the radio 
luminosities for a given intrinsic AGN power. To test this possibility, in Figures \ref{efficiency} and \ref{efficiency24} 
we present B$_{gq}^{av}$ versus 
the luminosity ratios L(5 GHz)/L([O III]$\lambda$5007) and L(5 GHz)/L(24 \micron) respectively. These ratios tell us how the radio 
luminosities of PRGs are affected by the environment for a given intrinsic AGN power, as indicated by the [O III] and 
24 \micron~luminosities \citep{Dicken09}.

From Figures \ref{efficiency} and \ref{efficiency24} we see that, below L(5 GHz)/L([O III]$\lambda$5007)$\sim$40 and 
L(5 GHz)/L(24 \micron)$\sim$0.1, there is no clear relationship with B$_{gq}$.  
The majority of SLRGs in the 2Jy sample are included in the previous limits. However, if we look at the sources with
the richest environments (B$_{gq}\ga800$; Abell class $>$1), they all have relatively large L(5 GHz)/L([O III]$\lambda$5007) and 
L(5 GHz)/L(24 \micron) ratios. Alternatively, all sources with L(5 GHz)/L([O III]$\lambda$5007)$\geq$100 and/or
L(5 GHz)/L(24 \micron)$\geq$0.3 reside in relatively rich environments (B$_{gq}\geq500$) and are WLRGs.  
The only exceptions are the SLRGs PKS 0409-75 and PKS 1306-09. 

Summarising, although we find no clear correlations between the environments of and the emission line luminosities and 
radio powers (see Figures \ref{lO3} and \ref{l5GHz}), we find 
that the objects with the largest clustering amplitudes --most of which are WLRGs/FRIs-- tend to
have larger ratios of radio luminosity to intrinsic AGN power. This might be the result of jet interactions
with a high density hot gas environment, which would likely favour a more efficient transformation of AGN power into 
radio luminosity. Alternatively, it could be a consequence of different accretion modes acting in WLRGs (i.e. cooling
flows or direct accretion of hot gas), or of the properties of the supermassive black holes (SMBHs) themselves, 
influenced by the environment. The merger
histories of central cluster galaxies may be different from those in the field, leading, for example, to more rapidly 
spinning black holes \citep{Fanidakis11} and to an increased incidence of radio-loud AGN.

\subsection{Comparison with control sample of quiescent early-type galaxies}
\label{controlsample}

In the previous sections we have discussed the role of environment on the triggering of PRGs and 
type-2 quasars, but it is necessary to compare these environments with those of the quiescent galaxies 
in the comparison samples, as we did with the galaxy interactions in \citealt{Ramos12} and \citet{Bessiere12}.

In order to do this, first, we
compared the environments of PRGs at $0.2\leq z\leq 0.7$ with the parent EGS control 
sample of 107 early-type galaxies. Second, for the type-2 quasars, we used the 51 early-type galaxies with 
redshifts $0.3\leq z\leq 0.41$ that comprise the EGS* control sample. 
In Tables \ref{Bgq_control_individual} and \ref{Bgq_control_individual2} we report the individual B$_{gq}$ values that we obtained
for the EGS and EGS* samples.

\begin{figure*}
\centering
{\par
\includegraphics[width=6.7cm]{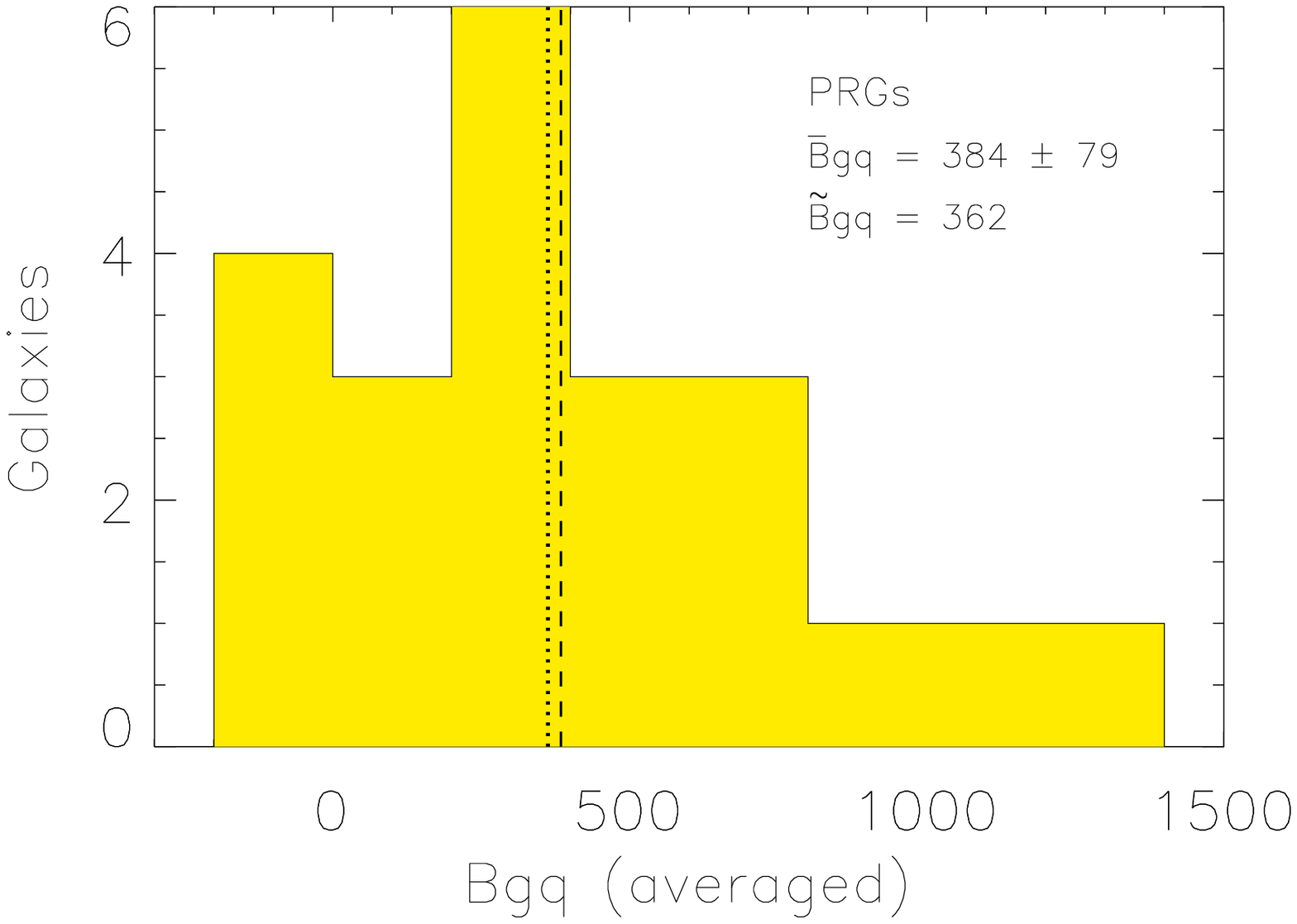}
\includegraphics[width=6.7cm]{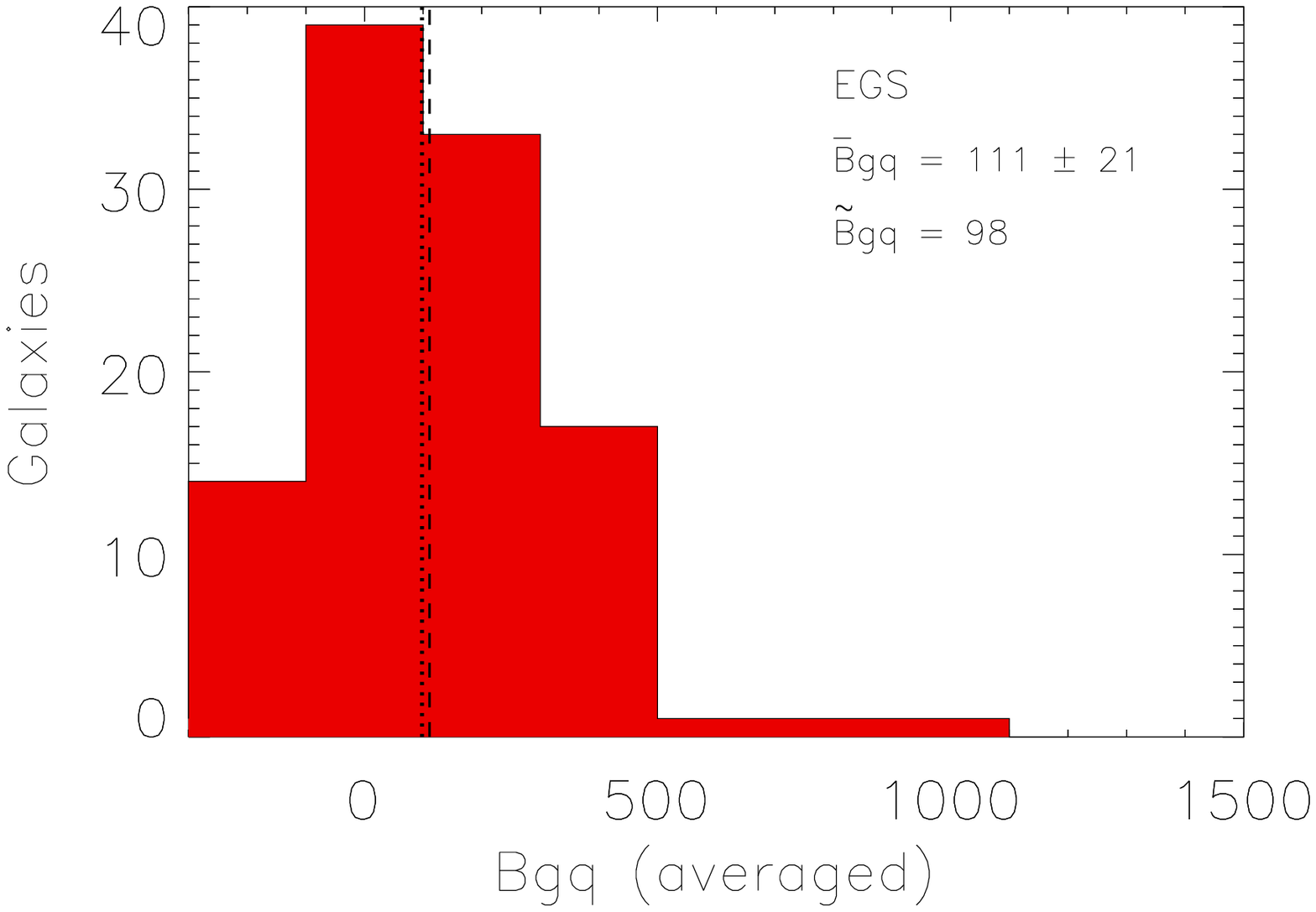}
\includegraphics[width=6.7cm]{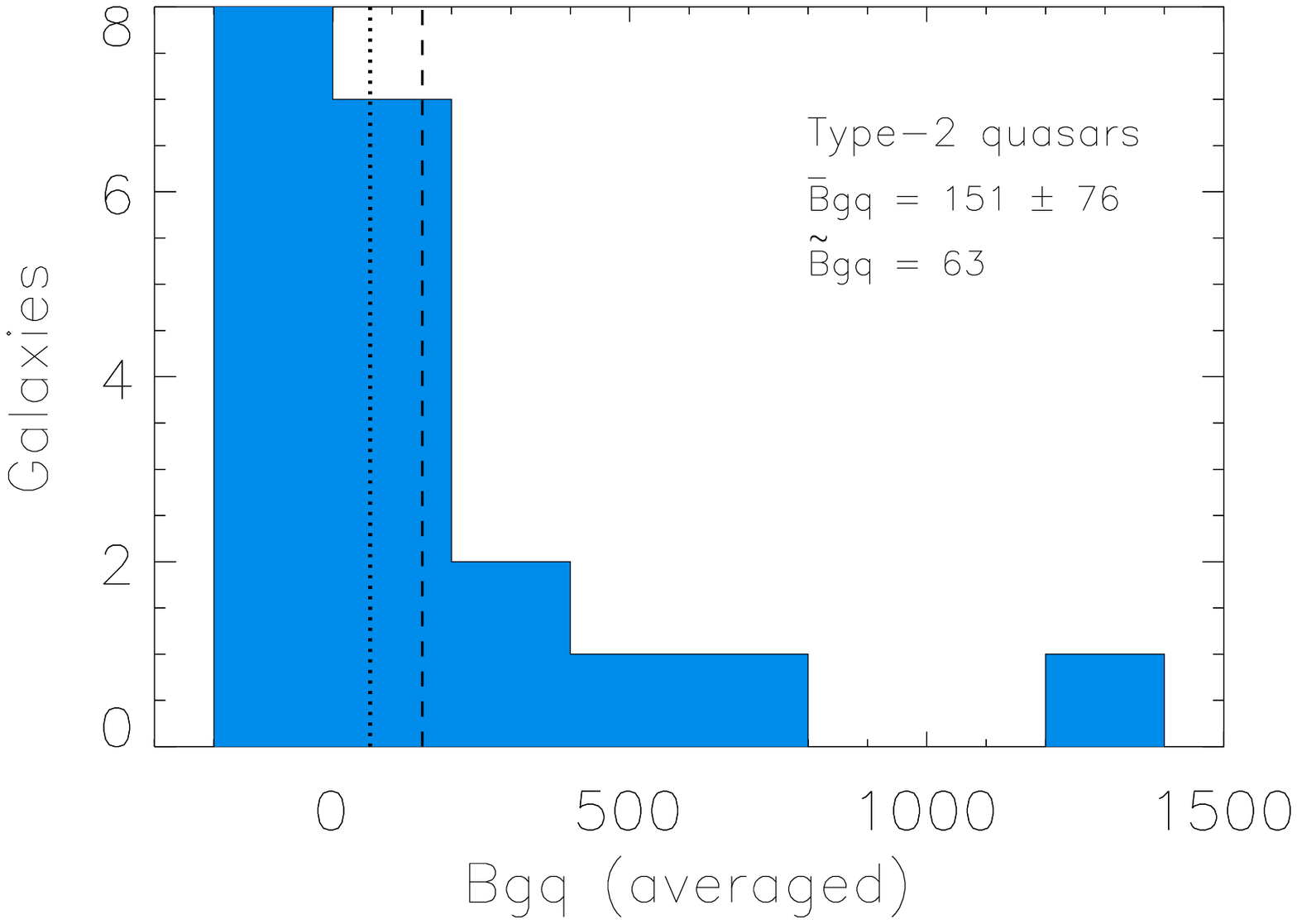}
\includegraphics[width=6.7cm]{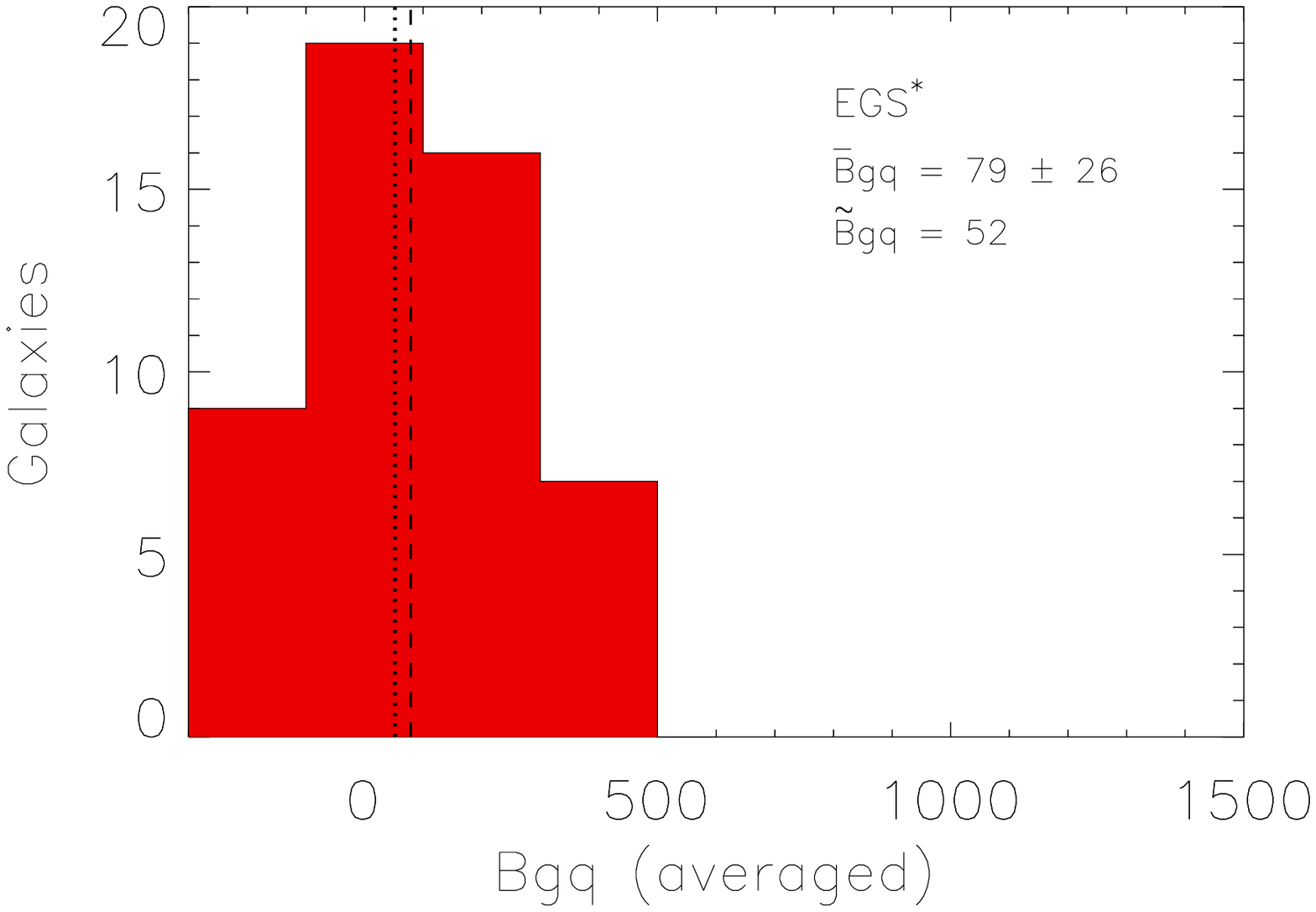}\par}
\caption{Distribution of spatial clustering amplitudes of the PRGs with redshifts $0.2\leq z<0.7$ and EGS galaxies (top panels)
and of the type-2 quasars and EGS* galaxies (bottom panels). 
Dotted and dashed lines correspond to the median and the average of the distribution respectively. Control sample histograms 
are represented as red (EGS and EGS*), PRGs as yellow and type-2 quasars as blue.}
\label{hist4}
\end{figure*}

In Table \ref{Bgq_control} we show the mean values of the distributions of B$_{gq}$, B$_{gq}^{av}$ and B$_{gq}^{med}$ 
that we measured for the EGS and EGS* samples, and the comparison with the PRGs and type-2 quasars. 
First, we find a significant difference 
between the environments of PRGs and EGS sample (see top panels of Figure \ref{hist4}). PRGs at $0.2\leq z<0.7$
(21 SLRGs and only one WLRG\footnote{This WLRG is PKS 0347+05, which is part of an interacting 
system together with a radio-quiet quasar \citep{Tadhunter12}.}) are, on average, in denser environments 
($\bar{B}_{gq}^{av}=384\pm79$) than their quiescent counterparts ($\bar{B}_{gq}^{av}=111\pm21$). This difference 
is significant at the 3$\sigma$ level according to the KS test (see Table \ref{Bgq_control}). 

Although we do not have a control sample suitable for the study of the environment 
for the low-redshift 2Jy PRGs ($z<0.2$; see Section \ref{quiescent}), the larger number of WLRGs 
in it (10 WLRGs and 14 SLRGs), as compared to the high-redshift subsample, 
increases the mean of B$_{gq}^{av}$ up to 450$\pm$91. Therefore, it seems logical to assume that
PRGs at $z<0.2$ also are in denser environments than quiescent early-type galaxies.



\begin{table}
\centering
\footnotesize
\begin{tabular}{lcccc}
\hline
\hline
Comparison    &  Targets & B$_{gq}$    & B$_{gq}^{av}$   &  B$_{gq}^{med}$ \\
\hline
PRGs	      &   22     & 344$\pm$85  & 384$\pm$79      &  389$\pm$79	\\ 	
EGS 	      &  107     & 112$\pm$20  & 111$\pm$21      &  101$\pm$21	\\ 
KS test       &  \dots   &  98.8\%     &  99.7\%         &   99.9\% 	\\ 
\hline		     			      		 
Type-2 quasars&  20      & 184$\pm$87  & 151$\pm$76      &  159$\pm$76	\\ 
EGS*          &  51      &  79$\pm$26  &  79$\pm$26      &   77$\pm$25	\\ 
KS test       &  \dots   &  21.8\%     &  15.5\%         &   40.4\% 	\\ 
\hline
\end{tabular}						 
\caption{Comparison between the cluster environments of 1) PRGs and whole EGS control sample ($0.2\leq z<0.7$)
and 2) type-2 quasars and EGS* control sample ($0.3\leq z\leq 0.41$). $\bar{B}_{gq}\pm\sigma(B_{gq})/\sqrt{n}$ is reported 
for each group, with n equal to the number of targets included in the mean, 
together with the results of the KS test (significance level).}
\label{Bgq_control}
\end{table}

The case of the type-2 quasars is different. We do not find a significant difference between the environments of 
active ($\bar{B}_{gq}^{av}=151\pm76$) and non-active early-type galaxies ($\bar{B}_{gq}^{av}=79\pm26$) 
at redshift $0.3\leq z\leq0.41$ and M$_B$=[-22.1,-20.3] mag (see Table 
\ref{Bgq_control} and bottom panels of Figure \ref{hist4}).

This result is in apparent contradiction with \citet{Serber06}, who claimed that, on scales ranging from 25 kpc to 1 Mpc, quasars
at $z\leq 0.4$  and M$_B$=[-23.0,-20.8] mag are located in denser environments than their quiescent counterparts. 
The counting radius that we are using (170 kpc) should be comparable to the small scales considered by 
\citet{Serber06}, but our analysis has the advantage of considering the galaxies contained in a volume rather than in an area.
On the other hand, on larger scales, of $\sim$1 Mpc, \citet{Serber06} found that quasars inhabit similar environments than the control 
sample galaxies.

As explained in the Introduction, the periods of black hole growth are coupled with the growth of the host galaxy. Consequently,
we do not expect to see a difference in the environment of radio-quiet quasars and quiescent early-type galaxies of the same 
mass and redshift. The AGN phase represents a very small fraction of the life of a massive galaxy, and the environment will 
not change significantly within the timespan of a single period of nuclear activity. 

In contrast, only some quiescent early-type galaxies have been/may be radio-loud AGN at some point, and the 
significant difference that we found between their environments and those of the control sample galaxies is 
noteworthy. It is not clear why only $\sim$10\% of the AGN population is radio-loud. The denser environments 
that we found here for radio-loud AGN, as compared to radio-quiet AGN and control sample galaxies, points out
a possible physical explanation behind the radio jet production. The high density hot gas environment characteristic
of clusters could be favouring the transformation of AGN power into radio luminosity. 
Alternatively, the properties of the SMBHs themselves could be influenced by the environment. The merger
histories of central cluster galaxies can lead to more rapidly 
spinning black holes \citep{Fanidakis11} and to an increased incidence of radio-loud AGN. 

The work presented here provides evidence for the radio-quiet AGN phase being an ubiquitous stage 
in the evolution of massive early-type galaxies, as well as for the environment to be responsible, to 
a certain extent, for the radio loudness of AGN\footnote{For a detailed study of the host galaxy properties of the 
2Jy sample we refer the reader to \citet{Inskip10} and Inskip et al.,~(in preparation).}. Our findings also support
the picture that radio-loud and radio-quiet AGN are independent phenomena.



In order to confirm the influence that the environment has in AGN radio power, 
it is important to use larger samples of PRGs, as well as to compare with X-ray information 
about the environment. In the future, we aim to repeat this study for the 3CRR sample of radio 
galaxies \citep{Laing83} and to compare the results found for the environment of the 2Jy PRGs with 
X-ray data from the XMM-Newton satellite (Mingo et al., in preparation).

\section{Conclusions}

We have presented the results from a quantitative analysis of the environments of complete samples of PRGs, type-2 quasars
and quiescent early-type galaxies. We have also investigated the connection between environment and the triggering 
mechanisms for nuclear activity in luminous radio-quiet and radio-loud AGN. Our major results are as follows:

\begin{itemize}

\item WLRGs in the 2Jy sample are in richer environments ($\bar{B}_{gq}^{av}=788\pm140$) than SLRGs 
($\bar{B}_{gq}^{av}=303\pm53$). This difference between their B$_{gq}$ distributions is significant at the 3$\sigma$ level, 
based on the KS test.
We obtain the same result when we compare the environment of FRI and FRII galaxies. 
WLRGs/FRIs have large ratios of radio luminosity versus AGN power--by definition--, constituting a first
indication that dense environments may boost the radio emission of PRGs. 

\item We do not observe an enhancement in the clustering of FRIIs with redshift. In fact, if we separate 
low-redhift FRIIs ($z<0.2$; 16 sources) and high-redshift FRIIs ($0.2\leq z<0.7$; 17 sources), we find similar
values of the spatial clustering amplitude: $\bar{B}_{gq}^{av}$($z<0.2$) = 351$\pm$98 and 
$\bar{B}_{gq}^{av}$($0.2\leq z<0.7$) = 340$\pm$90. 


\item When we compare the environments of type-2 quasars and PRGs in the 2Jy sample with [O III] luminosities larger
than 10$^{8.5} L_{\sun}$, we find that PRGs are more clustered than the type-2 quasars. However, this difference 
is significant at the 2$\sigma$ level only. A larger sample is 
required to put it on a firmer statistical footing.


\item If we consider the 20\% of the 2Jy sample with recent star formation activity detected, 
we find that 78\% of them (7 of the 9) are in clusters of Abell types 0, 1 and 2. Galaxy interactions 
could be leading to an enhancement of star formation in the galaxy hosts. Alternatively, 
cooling flows without a completely established AGN feedback could be favouring the formation of new stars. 

\item We do not find a significant difference between the environments of radio-quiet AGN 
and non-active early-type galaxies at redshift $0.3\leq z\leq0.41$ and M$_B$=[-22.1,-20.3] mag. This is consistent
with the quasar phase being a short-lived but ubiquitous stage in the formation of all massive
early-type galaxies. 

\item We find a significant difference (at the 3$\sigma$ level) between the environments of radio-loud AGN 
at $0.2\leq z<0.7$ ($\bar{B}_{gq}^{av}=384\pm79$) and their quiescent counterparts ($\bar{B}_{gq}^{av}=111\pm21$). 
This supports a physical origin for radio jet production, with high density hot gas environment favouring the 
transformation of AGN power into radio luminosity, or alternatively, with the environment influencing the properties
of the SMBHs themselves.   \\

\end{itemize}

\appendix

\section{Catalogue completeness.}
\label{appendixA}

The aim of this work is to compare the environments of PRGs, type-2 quasars and quiescent early-type galaxies and, based on 
that, discuss the role of environment on the triggering of nuclear activity. For these comparisons to be 
meaningful, the galaxy counts around each of the targets considered here have to be done to the same relative magnitude 
limit. Here we have used the criterion $(m_*-1)\leq m\leq (m_*+2)$ to count galaxies in both the target and the offset 
fields, and we need to show that the GMOS-S and Suprime-Cam data are deep enough to count galaxies down to the dimmest 
limit in each case $(m_*+2)$. 

As discussed in Section \ref{quiescent}, in \citealt{Ramos12}
we measured a median surface brightness of $\mu_V$=24.2~mag~arcsec$^{-2}$ for the tidal features detected in the 
galaxy hosts of the EGS galaxies, and a surface brightness range $22\leq\mu_V\leq26$~mag~arcsec$^{-2}$.
In addition, the seeing of the 4 Suprime-Cam images ranges from FWHM = 0.65\arcsec to 0.76\arcsec. 
Thus, the EGS and EGS* data are comparable in depth and resolution to the GMOS-S images employed in the study 
of PRGs and type-2 quasars. However, especially in the case of the 2Jy sample, the GMOS-S data span a wide
range of exposure times (from 250 s to 1500 s) and seeing FWHM (from 0.4\arcsec~to 1.15\arcsec). Thus, it becomes 
necessary to demonstrate that those images with large seeing values and/or low exposure times are sufficiently 
deep to count galaxies down to (m$^*$+2).

\begin{figure*}
\centering
\includegraphics[width=14cm]{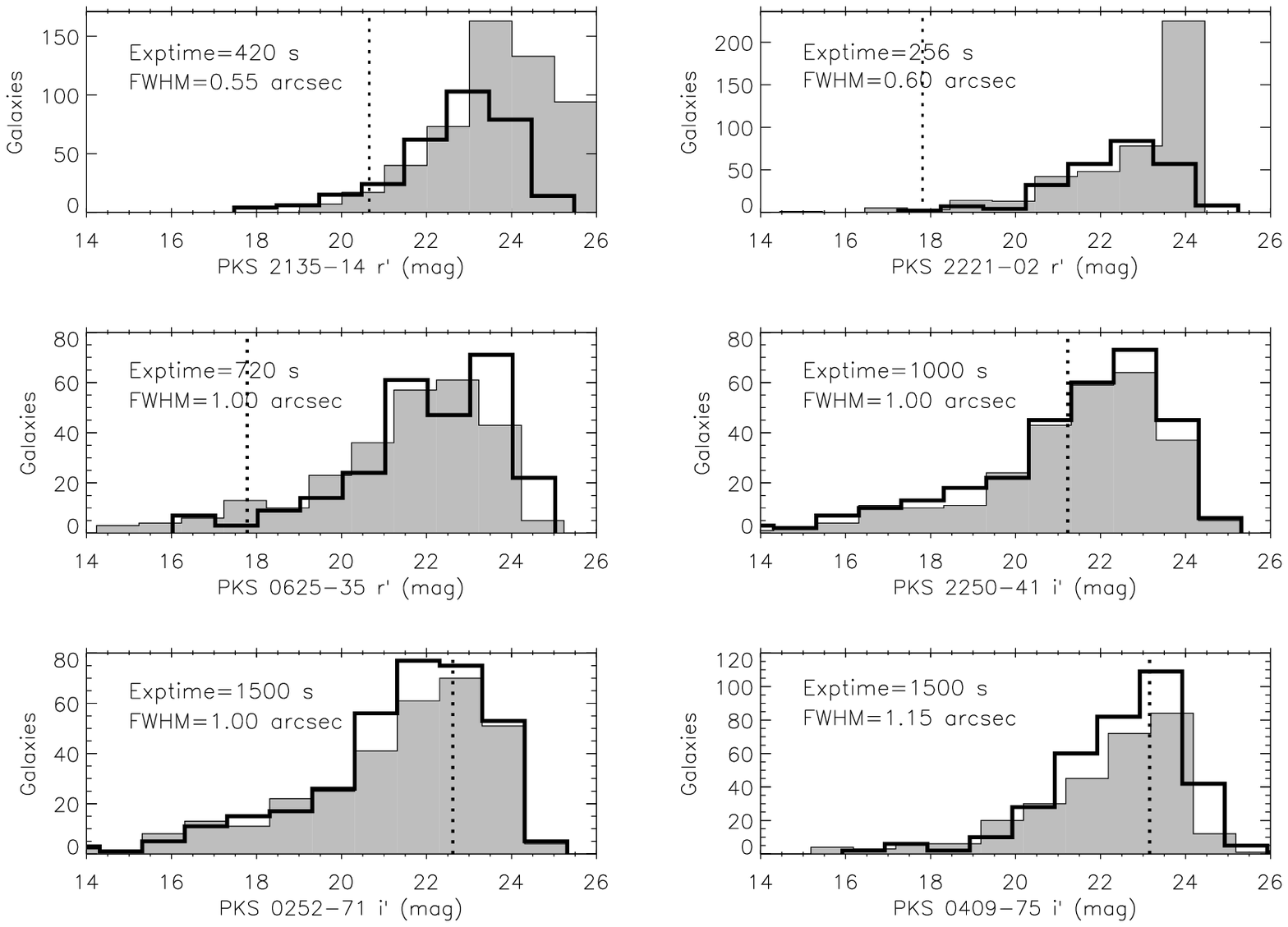}
\caption{Galaxy counts as a function of apparent magnitude (blank histograms) for the three galaxy 
fields with the lowest exposure times in the 2Jy and 
type-2 quasar samples (256, 420 and 720 s) and the three galaxy fields with the worst seeing (FWHM = 1.00-1.15 arcsec). 
Filled histograms represent the galaxy counts measured in the corresponding offset fields, which were observed
immediately after each target field with exposure times between 800 and 1500s. Vertical 
dotted lines indicate the position of the (m$^*$+2) limit used to count galaxies.}
\label{B1}
\end{figure*}

Figure \ref{B1} shows six blank histograms of galaxy counts as a function of apparent magnitude (in the r' and i'-bands)
for the three GMOS-S fields with the lowest exposure times (256, 420 and 720 s) and the three with the
worst seeing values (FWHM = 1.00-1.15\arcsec). The rest of 2Jy and type-2 quasars were observed with exposure times ranging 
between $\sim$1000 and 2000 s. We also included the galaxy counts measured in the corresponding offset fields (grey 
histograms), which were observed immediately after each target field, with exposure times ranging from 800 to 1500 s. 

\begin{figure*}
\centering
\includegraphics[width=14cm]{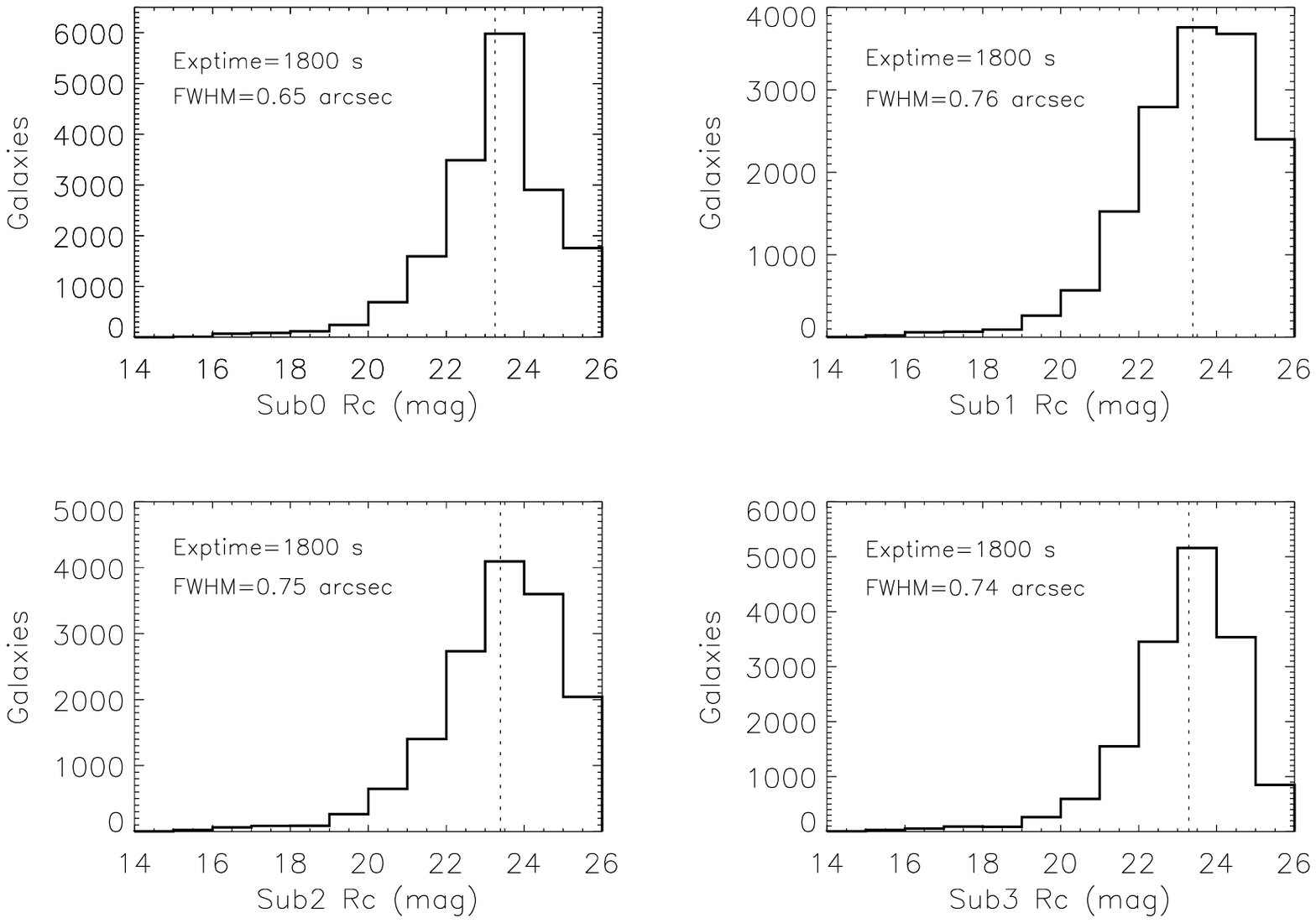}
\caption{Same as in Figure \ref{B1}, but for the four Suprime-Cam/Subaru fields from which the EGS and EGS* control sample
galaxies were taken from. Vertical 
dotted lines indicate the position of the faintest (m$^*$+2) limit used to count galaxies in each field.}
\label{B2}
\end{figure*}

\begin{figure*}
\centering
{\par
\includegraphics[width=7cm]{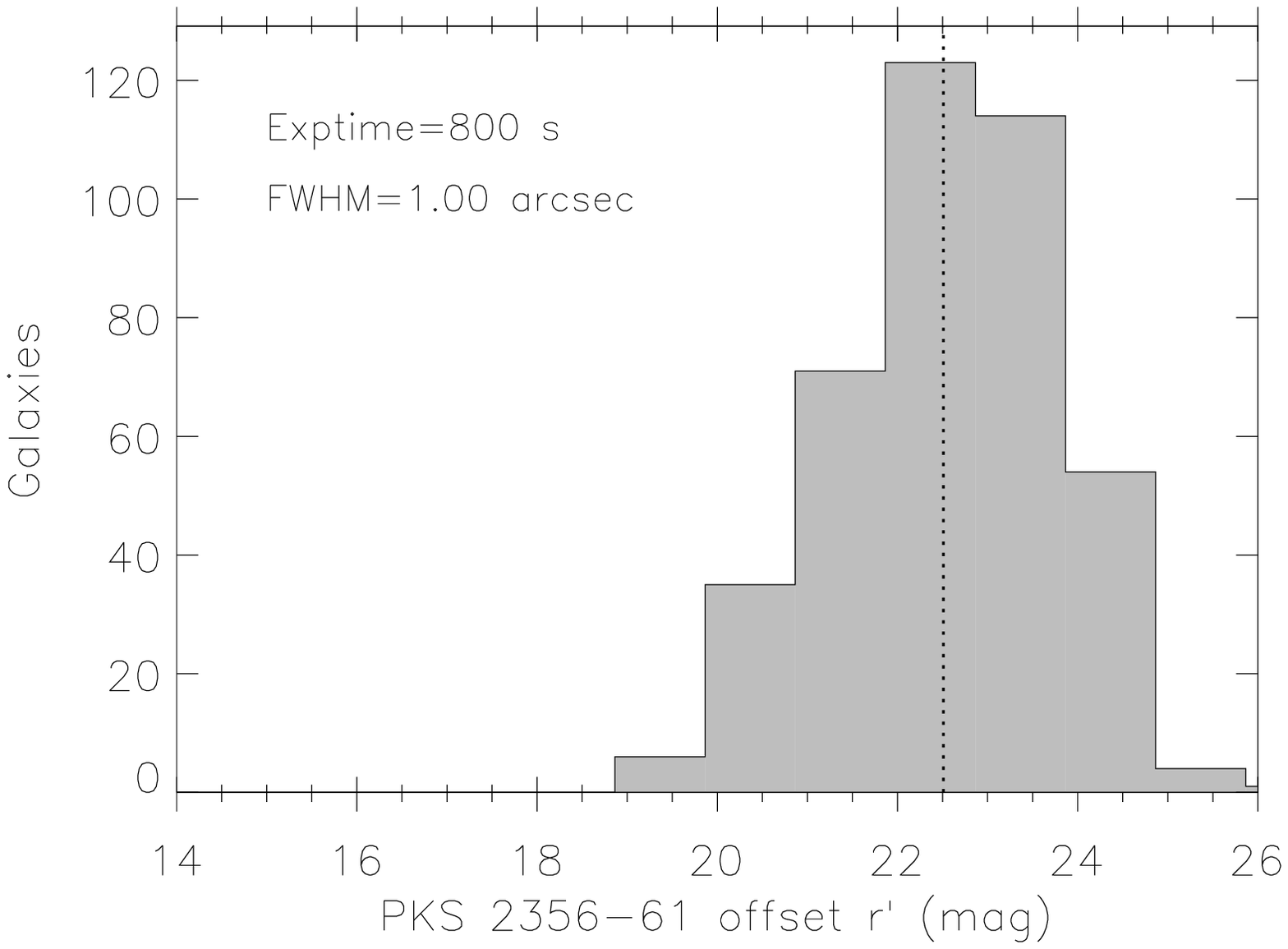}
\includegraphics[width=7cm]{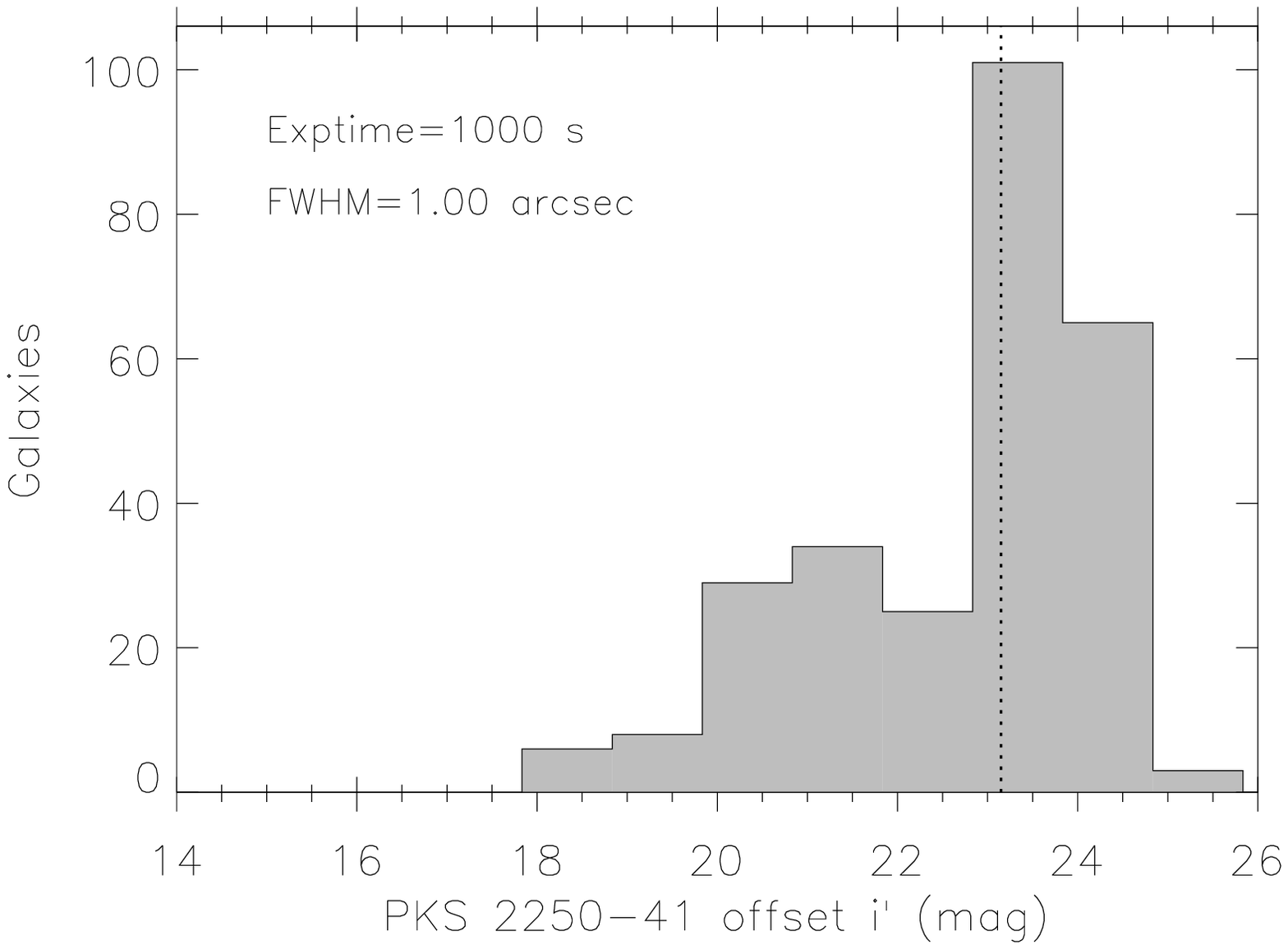}\par}
\caption{Galaxy counts as a function of apparent magnitude for the two offset fields with the lowest exposure times
and worse seeing FWHMs in the r' and i'-bands respectively. Vertical 
dotted lines indicate the position of the faintest (m$^*$+2) limits among all the targets considered in this work.}
\label{B3}
\end{figure*}

The histograms in Figure \ref{B1} show a maximum around 23.5 mag, and a sharp 
cut at $\sim$25 mag. 
The same behaviour is shown by the galaxy counts in the four Suprime-Cam EGS fields
(see Figure \ref{B2}). In this case, all the fields were observed with the same exposure time and under similar seeing 
conditions (1800 s and $\sim$0.70\arcsec). The larger galaxy counts are due to the different field sizes
(34$\times$27 arcmin$^2$ for Suprime-Cam and 5.5$\times$5.5 arcmin$^2$ for GMOS-S).
Finally, in Figure \ref{B3} we show the number of counts measured in the two offset fields with the lowest exposure times
and worse seeing FWHMs in the r' and i'-bands respectively.

The histograms in Figures \ref{B1}, \ref{B2} and \ref{B3} were plotted after discarding stars, sources close 
to image boundaries and with saturated and/or corrupted pixels using the CLASS$_-$STAR and FLAG SExtractor 
parameters, as described in Section \ref{catalogs}.
By comparing Figures \ref{B1}, \ref{B2} and \ref{B3} it is clear that both the control sample and offset field 
images are comparable in depth and resolution to the PRG and type-2 quasar images. 

The vertical dotted lines in Figure \ref{B1}
correspond to the (m$^*$+2) limit for each target, which basically depends on the galaxy redshift. In Figures \ref{B2} and \ref{B3}, 
the vertical lines indicate the faintest (m$^*$+2) limit in each of the four Subaru fields and among all the target fields respectively. 
Even for the fields with the worst quality data (i.e. largest seeing FWHMs and lowest exposure times), 
the (m$^*$+2) limits are brighter or equal to the maximum of galaxy count distributions. Note that the last two histograms in Figure
\ref{B1} correspond to galaxies with the highest redshifts in the 2Jy and type-2 quasar samples (z=0.6 and 0.7 respectively), 
and thus, with the dimmest (m$^*$+2) limit. Therefore, we can confidently compare the clustering amplitudes of PRGs, type-2 quasars
and control sample galaxies obtained in this work without running into completeness issues.

\section{Galaxy counts and luminosity function normalization.}
\label{appendixB}

\begin{figure*}
\centering
{\par
\includegraphics[width=3.8cm,angle=90]{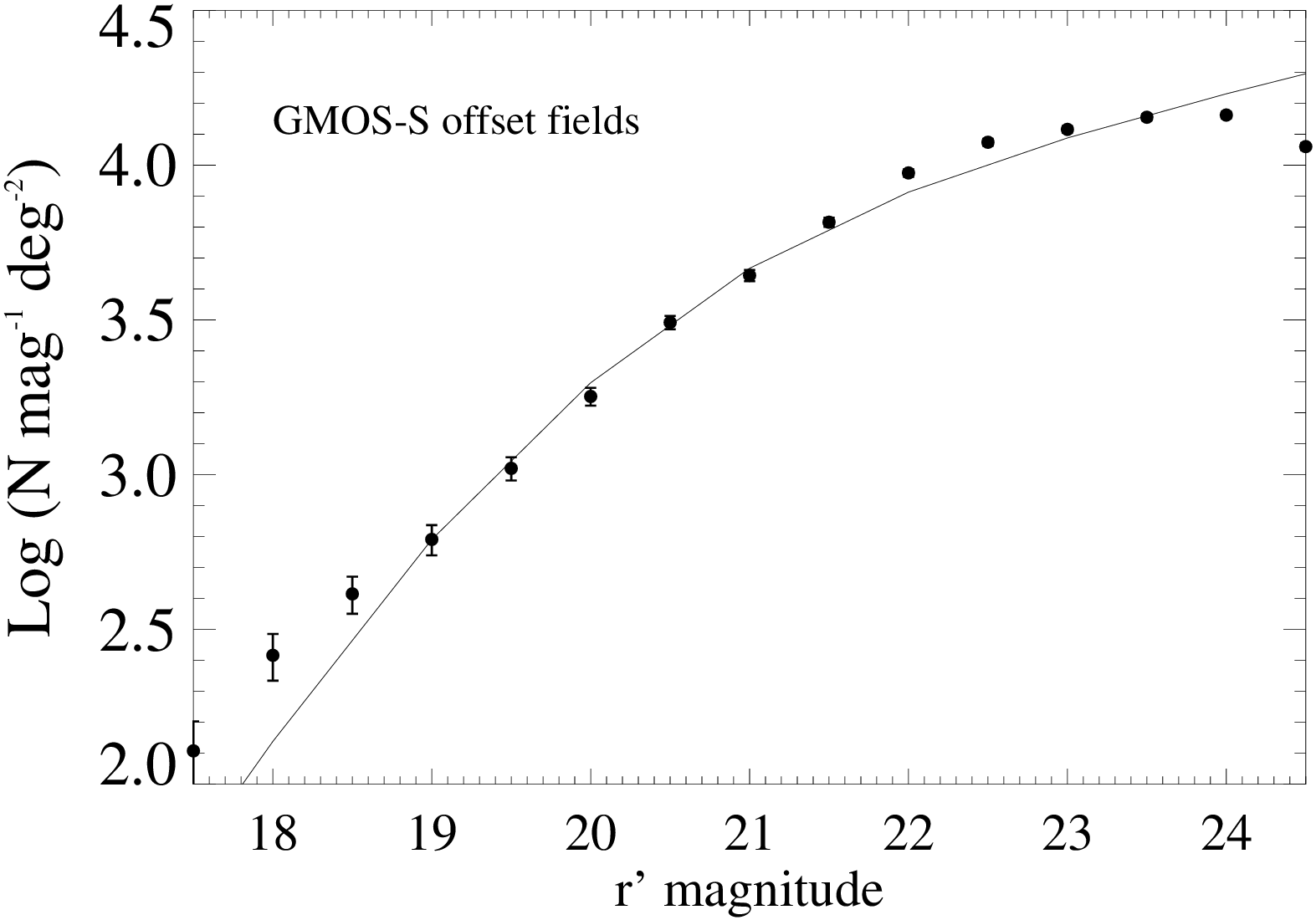}
\includegraphics[width=3.8cm,angle=90]{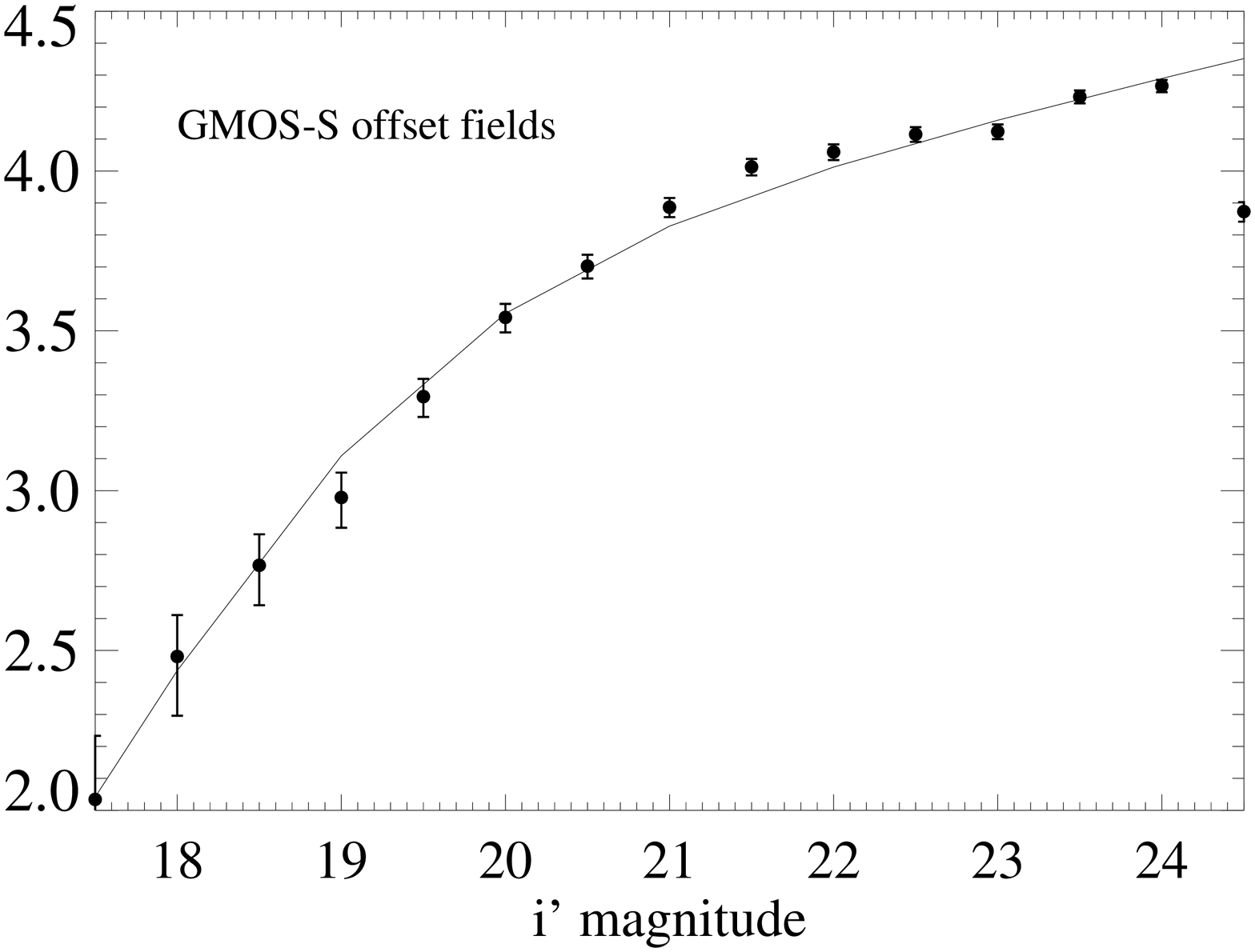}
\includegraphics[width=3.8cm,angle=90]{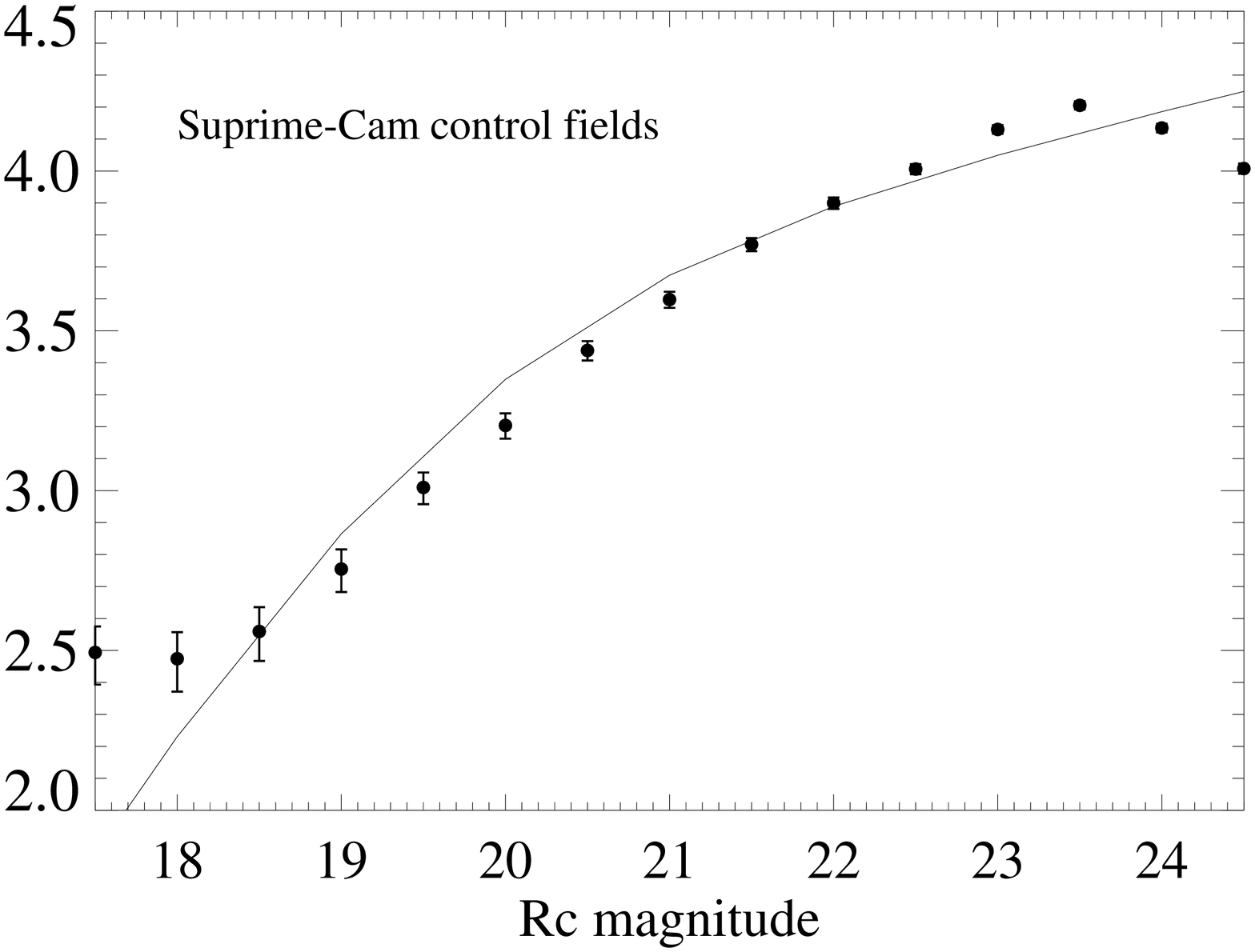}\par}
\caption{Background counts, represented as black dots, measured from the GMOS-S r' and i'-band offset fields (left and central panels) 
and in the four Suprime-Cam Rc images (right panel). Solid lines are the predicted background counts, obtained from integrating the 
evolving luminosity function along the line of sight considering the five redshift bins indicated in Table \ref{Schechter}. }
\label{C1}
\end{figure*}

As described in Section \ref{cluster}, the clustering amplitudes discussed in this work depend on the chosen luminosity function. 
To demonstrate that the luminosity function parameters given in Table \ref{Schechter} are consistent with our background
counts, we integrated our evolving luminosity function along the line of sight considering the five redshift bins indicated
in Table \ref{Schechter}. The predicted background counts as a function of apparent magnitude in the GMOS-S r' and i'-band 
filters and in the Suprime-Cam Rc filter are shown as solid lines in the three panels of Figure \ref{C1}. Using our 
galaxy catalogs, we counted galaxies after getting rid of stars and sources close to image boundaries, 
or with saturated and/or corrupted pixels 
(using the CLASS$_-$STAR and FLAG SExtractor parameters as described in Section \ref{catalogs}). We computed 
the average background counts in the 52 GMOS-S r'-band offset fields, in the 11 i'-band offset fields, and in 
the four Suprime-Cam Rc images (black dots in Figure \ref{C1}). We calculated Poissonian errors multiplied by a 
1.3 factor to approximate possible departures from Poisson statistics \citep{Yee99,Wold00}.

Figure \ref{C1} shows that our choice of evolving luminosity function is consistent with the data. The agreement 
between the predicted and measured number counts and the faint end, up to a limit of 23-24 mag, backs up the results
of Appendix \ref{appendixA}. 

In the case of the Rc and r'-band fields (left and right panels of Figure \ref{C1}), we 
detect an excess in the number of background counts at magnitudes brighter than 19. After visual inspection of the 
individual images, we found out that this excess is due to intruder stars that have not been removed from our catalogues. 
For example, among the 52 r'-band fields, there are 33 sources with $0.7\leq STAR\leq0.85$ and r'$\leq$19 mag which are
either stars with small deviations from symmetry produced by extremelly faint galaxies next to them, 
or stars immersed in the bright haloes of saturated stars. These 33 sources are distributed in 8 out of the 52 GMOS-S r'-band
fields. This explains the lack of bright excess in the i'-band counts (central panel of Figure \ref{C1}), which were
measured in 11 offset fields only. In the Rc-band Suprime-Cam fields we see the same effect as in the GMOS-S r'-band,
because we are measuring galaxy counts in an even larger area ($\sim$1 deg$^2$).

\section{Clustering amplitudes of control sample galaxies.}
\label{appendixC}

Here we present the individual spatial clustering amplitudes of the 107 early-type galaxies in the EGS sample (Table
\ref{Bgq_control_individual}) and 51 early-type galaxies in the EGS* sample (Table \ref{Bgq_control_individual2}). 

\begin{table*}
\centering
\begin{tabular}{lcccccccccl}
\hline
\hline
irac ID      & z$_{PHOT}$ &  N$_t$ & N$_b$ &  B$_{gq}$ &  N$_b^{av}$   & $\sigma$  &   B$_{gq}^{av}\pm\Delta B_{gq}^{av}$ & N$_b^{med}$ &  B$_{gq}^{med}$ & Morphology	  \\
 (1)         &  (2)       & (3)    & (4)   & (5)       & (6)           &   (7)     &  (8)	     & (9)	   & (10)	     & (11)	 \\  
\hline
      004162 & 0.48  &     5   &    1.97   &   351   &     2.08   &   0.13  &    338$\pm$294   &    2.18   &	   326   & ...  	 \\
      006612 & 0.31  &     6   &    1.30   &   447   &     1.42   &   0.08  &    436$\pm$252   &    1.45   &	   433   & B,F,[D],[T	 \\
      006613 & 0.30  &     6   &    1.26   &   446   &     1.37   &   0.08  &    436$\pm$248   &    1.42   &	   431   & B		 \\
 056690$_-$1 & 0.50  &     4   &    2.32   &   199   &     2.17   &   0.15  &    217$\pm$278   &    2.27   &	   205   & [A],[B]       \\
      060191 & 0.57  &     5   &    2.60   &   307   &     2.46   &   0.21  &    324$\pm$330   &    2.60   &  	   307   & F	 \\
      060958 & 0.40  &     3   &    1.75   &   132   &     1.74   &   0.08  &    133$\pm$216   &    1.75   &	   132   & T,[A],[B]	 \\
      061249 & 0.65  &     4   &    3.10   &   125   &     2.88   &   0.32  &    154$\pm$336   &    3.10   &	   125   & [T]  	 \\
      066105 & 0.51  &     4   &    2.31   &   202   &     2.20   &   0.16  &    215$\pm$281   &    2.31   &	   202   & [A]  	 \\
      067417 & 0.39  &     0   &    1.71   &  -178   &     1.71   &   0.07  &   -178$\pm$113   &    1.71   &      -178   & ...		 \\
      072533 & 0.33  &     1   &    1.49   &   -47   &     1.49   &   0.07  &    -47$\pm$137   &    1.49   &	   -48   & S		 \\
      073519 & 0.49  &     0   &    2.21   &  -259   &     2.11   &   0.14  &   -247$\pm$141   &    2.21   &      -259   & [A]		 \\
      074777 & 0.42  &     6   &    1.86   &   449   &     1.82   &   0.07  &    453$\pm$292   &    1.86   &	   449   & [S]  	 \\
      074924 & 0.41  &     5   &    1.79   &   344   &     1.77   &   0.07  &    345$\pm$266   &    1.79   &	   344   & ...  	 \\
      077695 & 0.35  &     1   &    1.58   &   -58   &     1.57   &   0.08  &    -57$\pm$144   &    1.58   &	   -58   & T		 \\
      079968 & 0.60  &     5   &    2.80   &   290   &     2.62   &   0.25  &    313$\pm$343   &    2.80   &	   290   & F		 \\
      082325 & 0.55  &     2   &    2.51   &   -63   &     2.38   &   0.20  &    -47$\pm$236   &    2.51   &	   -63   & [F]  	 \\
      083714 & 0.50  &     5   &    2.27   &   323   &     2.17   &   0.15  &    335$\pm$302   &    2.27   &	   323   & F		 \\
      088031 & 0.50  &     4   &    2.03   &   233   &     2.17   &   0.15  &    217$\pm$278   &    2.27   &	   205   & F		\\
      090430 & 0.38  &     3   &    1.67   &   137   &     1.68   &   0.07  &    137$\pm$212   &    1.69   &	   136   & A,F,[B]	\\
      092065 & 0.55  &     3   &    2.19   &   101   &     2.38   &   0.20  &     78$\pm$271   &    2.51   &	    61   & B		\\
      092765 & 0.35  &     2   &    1.55   & 	44   &     1.57   &   0.08  &     42$\pm$172   &    1.58   &	    41   & [A],[T]	\\
 093764$_-$1 & 0.39  &     2   &    1.70   & 	31   &     1.71   &   0.07  &     30$\pm$184   &    1.71   &	    30   & [S]  	\\
      094231 & 0.41  &     2   &    1.75   & 	27   &     1.77   &   0.07  &     24$\pm$187   &    1.79   &	    22   & 2F,[T]	\\
      094966 & 0.46  &     1   &    1.91   &  -103   &     2.00   &   0.11  &   -113$\pm$174   &    2.09   &      -123   & 2T	\\
      095727 & 0.38  &     3   &    1.67   &   137   &     1.68   &   0.07  &    137$\pm$212   &    1.69   &	   136   & F,S  	\\
      099954 & 0.27  &     1   &    1.33   &   -29   &     1.27   &   0.10  &    -24$\pm$122   &    1.33   &	   -29   & [T]  	\\
      102757 & 0.22  &     3   &    1.27   &   147   &     1.19   &   0.09  &    154$\pm$166   &    1.21   &	   152   & 2S		\\
      102982 & 0.60  &     4   &    2.42   &   208   &     2.62   &   0.25  &    181$\pm$316   &    2.80   &	   158   & F		\\
      103198 & 0.38  &     1   &    1.67   &   -69   &     1.68   &   0.07  &    -70$\pm$151   &    1.69   &	   -71   & 2N,F,S	\\
      104038 & 0.46  &     4   &    1.91   &   236   &     2.00   &   0.11  &    226$\pm$262   &    2.09   &	   216   & B		\\
      104729 & 0.63  &     1   &    2.52   &  -207   &     2.77   &   0.29  &   -241$\pm$232   &    2.97   &      -268   & A	\\
      105193 & 0.23  &     2   &    1.26   & 	63   &     1.19   &   0.08  &     69$\pm$143   &    1.23   &	    66   & [S]  	\\
      106324 & 0.26  &     3   &    1.33   &   149   &     1.25   &   0.10  &    156$\pm$175   &    1.33   &	   149   & [T]  	\\
      106984 & 0.45  &     3   &    1.88   &   125   &     1.95   &   0.10  &    117$\pm$232   &    2.02   &	   110   & A,[I]	\\
      111427 & 0.32  &     2   &    1.47   & 	51   &     1.45   &   0.09  &     52$\pm$164   &    1.47   &	    51   & 2N,T,2I	\\
      112580 & 0.51  &     4   &    2.06   &   232   &     2.20   &   0.16  &    215$\pm$281   &    2.31   &	   202   & [B]  	\\
      113088 & 0.48  &     3   &    1.97   &   119   &     2.08   &   0.13  &    106$\pm$243   &    2.18   &	    94   & [B]  	\\
      113577 & 0.67  &     2   &    2.73   &  -102   &     3.01   &   0.35  &   -143$\pm$286   &    3.24   &      -176   & [A]		\\
      114966 & 0.61  &     7   &    2.45   &   606   &     2.67   &   0.27  &    578$\pm$397   &    2.84   &	   554   & 2T,S 	\\
      115327 & 0.35  &     2   &    1.55   & 	44   &     1.57   &   0.08  &     42$\pm$172   &    1.58   &	    41   & 2F,[T]	\\
      115594 & 0.31  &     2   &    1.45   & 	52   &     1.42   &   0.08  &     55$\pm$164   &    1.45   &	    52   & 2N,T 	\\
      118942 & 0.37  &     1   &    1.64   &   -65   &     1.63   &   0.06  &    -64$\pm$148   &    1.65   &	   -66   & ...  	\\
      119696 & 0.50  &     2   &    2.03   & 	-3   &     2.17   &   0.15  &    -19$\pm$209   &    2.27   &	   -32   & B,F  	\\
      122098 & 0.22  &     0   &    1.27   &  -108   &     1.19   &   0.09  &   -101$\pm$77    &    1.21   &      -102   & ...		\\
      124509 & 0.34  &     4   &    1.52   &   244   &     1.52   &   0.07  &    244$\pm$221   &    1.53   &	   243   & B,2T,F	\\
      125663 & 0.53  &     2   &    2.13   &   -15   &     2.29   &   0.18  &    -35$\pm$228   &    2.41   &	   -50   & [F]  	\\
 126918$_-$1 & 0.49  &     4   &    2.00   &   234   &     2.11   &   0.14  &    221$\pm$273   &    2.21   &	   209   & F,[B]	\\
      127241 & 0.59  &     4   &    2.36   &   213   &     2.56   &   0.24  &    187$\pm$312   &    2.73   &	   165   & ...  	\\
      127457 & 0.50  &     2   &    2.03   & 	-3   &     2.17   &   0.15  &    -19$\pm$209   &    2.27   &	   -32   & 2N,A 	\\
      128074 & 0.34  &     4   &    1.52   &   244   &     1.52   &   0.07  &    244$\pm$221   &    1.53   &	   243   & B,[F]	\\
      128416 & 0.58  &     6   &    2.33   &   474   &     2.52   &   0.23  &    449$\pm$359   &    2.67   &	   430   & ...  	\\
      132682 & 0.33  &     1   &    1.40   &   -39   &     1.49   &   0.07  &    -47$\pm$137   &    1.49   &	   -48   & ...  	\\
      135859 & 0.40  &     2   &    1.66   & 	36   &     1.74   &   0.08  &     27$\pm$186   &    1.75   &	    26   & [I]  	\\
\hline		          				  
\end{tabular}	          				  
\end{table*}

\begin{table*}
\centering
\begin{tabular}{lcccccccccl}
\hline
\hline
irac ID      & z$_{PHOT}$ &  N$_t$ & N$_b$ &  B$_{gq}$ &  N$_b^{av}$   & $\sigma$  &   B$_{gq}^{av}\pm\Delta B_{gq}^{av}$ & N$_b^{med}$ &  B$_{gq}^{med}$ & Morphology	  \\
 (1)         &  (2)       & (3)    & (4)   & (5)       & (6)           &   (7)     &  (8)	     &  (9)	   & (10)	     & (11)	 \\  
\hline
      138794 & 0.50  &      3	 &     2.05  &     113    &  	2.17  &     0.15   &     98$\pm$250    &     2.27  &	  86	& [T]	   \\
      139190 & 0.44  &      0	 &     1.82  &    -201    &  	1.91  &     0.09   &   -211$\pm$127    &     1.97  &	-218	& ...	   \\
      140456 & 0.30  &      5	 &     1.26  &     352    &  	1.37  &     0.08   &    342$\pm$230    &     1.42  &	 336	& 2T	   \\
      140758 & 0.43  &      1	 &     1.78  &     -85    &  	1.87  &     0.08   &    -94$\pm$164    &     1.92  &	-101	& S	   \\
      141714 & 0.44  &      4	 &     1.82  &     242    &  	1.91  &     0.09   &    232$\pm$256    &     1.97  &	 225	& [B],[S]  \\
      143149 & 0.37  &      3	 &     1.55  &     148    &  	1.63  &     0.06   &    139$\pm$206    &     1.65  &	 138	& T	   \\
      143536 & 0.50  &      1	 &     2.05  &    -123    &  	2.17  &     0.15   &   -138$\pm$186    &     2.27  &	-150	& [T]	   \\
      145098 & 0.32  &      3	 &     1.35  &     159    &  	1.45  &     0.09   &    149$\pm$192    &     1.47  &	 147	& A,T	   \\
      145434 & 0.48  &      2	 &     1.96  &       4    &  	2.08  &     0.13   &     -9$\pm$209    &     2.18  &	 -21	& 4T	   \\
      146298 & 0.59  &      2	 &     2.34  &     -45    &  	2.56  &     0.24   &    -73$\pm$253    &     2.73  &	 -95	& [A]	   \\
      152722 & 0.49  &      2	 &     1.99  &       1    &  	2.11  &     0.14   &    -13$\pm$220    &     2.21  &	 -25	& [F]	   \\
      156161 & 0.30  &      3	 &     1.26  &     164    &  	1.37  &     0.08   &    153$\pm$186    &     1.42  &	 148	& T	   \\
      157751 & 0.47  &      2	 &     1.94  &       6    &  	2.05  &     0.12   &     -5$\pm$185    &     2.14  &	 -16	& ...	   \\
      157878 & 0.46  &      3	 &     1.90  &     125    &  	2.00  &     0.11   &    113$\pm$236    &     2.09  &	 103	& F	   \\
      159123 & 0.56  &      0	 &     2.27  &    -286    &  	2.43  &     0.21   &   -307$\pm$164    &     2.56  &	-324	& T	   \\
      159936 & 0.41  &      1	 &     1.70  &     -75    &  	1.77  &     0.07   &    -83$\pm$161    &     1.79  &	 -84	& 2N	   \\
      160442 & 0.47  &      5	 &     1.94  &     350    &  	2.05  &     0.12   &    338$\pm$290    &     2.14  &	 327	& B,A	   \\
      160500 & 0.34  &      5	 &     1.43  &     352    &  	1.52  &     0.07   &    343$\pm$242    &     1.53  &	 342	& B,2T     \\	
      161724 & 0.34  &      4	 &     1.43  &     253    &  	1.52  &     0.07   &    244$\pm$221    &     1.53  &	 243	& [F]	   \\
      165265 & 0.67  &      4	 &     2.69  &     185    &  	3.01  &     0.35   &    140$\pm$349    &     3.24  &	 107	& B,T	   \\
      166730 & 0.36  &      1	 &     1.51  &     -51    &  	1.60  &     0.07   &    -61$\pm$147    &     1.62  &	 -62	& S,T	   \\
      169386 & 0.47  &      0	 &     1.94  &    -222    &  	2.05  &     0.12   &   -234$\pm$136    &     2.14  &	-245	& ...	   \\
      172474 & 0.51  &      2	 &     2.07  &      -8    &  	2.20  &     0.16   &    -24$\pm$225    &     2.31  &	 -37	& T,B,F    \\
      173901 & 0.32  &      0	 &     1.35  &    -129    &  	1.45  &     0.09   &   -140$\pm$97     &     1.47  &	-141	& ...	   \\
      175347 & 0.60  &      1	 &     2.39  &    -184    &  	2.62  &     0.25   &   -214$\pm$221    &     2.80  &	-236	& S,[B]    \\
      175590 & 0.56  &      1	 &     2.65  &    -209    &  	2.43  &     0.21   &   -180$\pm$206    &     2.56  &	-197	& [A]	   \\
      177990 & 0.25  &      1	 &     1.11  &      -9    &  	1.23  &     0.10   &    -20$\pm$118    &     1.30  &	 -26	& F,[2N]   \\
      178118 & 0.46  &      5	 &     2.11  &     328    &  	2.00  &     0.11   &    340$\pm$286    &     2.09  &	 330	& ...	   \\
      178724 & 0.52  &      3	 &     2.43  &      69    &  	2.25  &     0.17   &     91$\pm$259    &     2.36  &	  77	& A	   \\
      178868 & 0.37  &      4	 &     1.70  &     235    &  	1.63  &     0.06   &    242$\pm$231    &     1.65  &	 240	& F	   \\
      180420 & 0.54  &      9	 &     2.54  &     800    &  	2.33  &     0.19   &    825$\pm$403    &     2.46  &	 809	& 2N,2T,[B]\\
      181402 & 0.38  &      6	 &     1.75  &     439    &  	1.68  &     0.07   &    447$\pm$277    &     1.69  &	 446	& [I],[A]  \\
      181444 & 0.31  &      1	 &     1.50  &     -47    &  	1.42  &     0.08   &    -39$\pm$131    &     1.45  &	 -42	& 2S,[I]   \\
      181736 & 0.46  &      6	 &     2.11  &     441    &  	2.00  &     0.11   &    453$\pm$308    &     2.09  &	 443	& ...	   \\
      181914 & 0.36  &      0	 &     1.69  &    -170    &  	1.60  &     0.07   &   -162$\pm$106    &     1.62  &	-163	& ...	   \\
      182762 & 0.43  &      4	 &     1.94  &     226    &  	1.87  &     0.08   &    234$\pm$253    &     1.92  &	 227	& [F]	   \\
      183081 & 0.36  &      6	 &     1.69  &     436    &  	1.60  &     0.07   &    444$\pm$269    &     1.62  &	 443	& F,[T]    \\
      183836 & 0.44  &      3	 &     2.00  &     111    &  	1.91  &     0.09   &    121$\pm$231    &     1.97  &	 114	& [S]	   \\
      184041 & 0.53  &      4	 &     2.48  &     186    &  	2.29  &     0.18   &    209$\pm$289    &     2.41  &	 194	& F,S	   \\
      184315 & 0.50  &      3	 &     2.32  &      80    &  	2.17  &     0.15   &     98$\pm$250    &     2.27  &	  86	& 2N	   \\
      186058 & 0.54  &      3	 &     2.54  &      57    &  	2.33  &     0.19   &     82$\pm$263    &     2.46  &	  66	& [A]	   \\
      189727 & 0.64  &      4	 &     3.15  &     116    &  	2.83  &     0.31   &    161$\pm$336    &     3.03  &	 133	& ...	   \\
      190795 & 0.51  &      4	 &     2.37  &     195    &  	2.20  &     0.16   &    215$\pm$281    &     2.31  &	 202	& T,S	   \\
      193464 & 0.42  &      3	 &     1.90  &     119    &  	1.82  &     0.07   &    128$\pm$224    &     1.86  &	 124	& 2N,F     \\
      193507 & 0.47  &      1	 &     2.16  &     899    &  	2.05  &     0.12   &    912$\pm$388    &     2.14  &	 901	& 2N,[B]   \\
      193737 & 0.50  &      2	 &     2.32  &     -37    &  	2.17  &     0.15   &    -19$\pm$209    &     2.27  &	 -32	& ...	   \\
      193974 & 0.40  &      2	 &     1.84  &      17    &  	1.74  &     0.08   &     27$\pm$186    &     1.75  &	  26	& [S]	   \\
      194092 & 0.51  &      5	 &     2.37  &     314    &  	2.20  &     0.16   &    335$\pm$305    &     2.31  &	 321	& [T]	   \\
      196827 & 0.37  &      4	 &     1.70  &     235    &  	1.63  &     0.06   &    242$\pm$231    &     1.65  &	 240	& T	   \\	
      198295 & 0.54  &      2	 &     2.54  &     -66    &  	2.33  &     0.19   &    -41$\pm$236    &     2.46  &	 -57	& [S]	   \\
      199503 & 0.50  &      1	 &     2.32  &    -156    &  	2.17  &     0.15   &   -138$\pm$186    &     2.27  &	-150	& T	   \\
      202111 & 0.27  &      1	 &     1.37  &     -33    &  	1.27  &     0.10   &    -24$\pm$122    &     1.33  &	 -29	& [S]	   \\
      204161 & 0.62  &      2	 &     3.01  &    -136    &  	2.72  &     0.28   &    -97$\pm$265    &     2.92  &	-123	& A,[B]    \\
      204944 & 0.28  &      2	 &     1.39  &      55    &  	1.31  &     0.09   &     63$\pm$156    &     1.36  &	  58	& T,S	   \\
\hline		     			      		 
\end{tabular}						 
\caption{Same as in Table \ref{Bgq_individual} but for the 107 early-type galaxies in the EGS sample. Last column corresponds to the
morphological classification in \citealt{Ramos12}: T: Tail; F: Fan; B: Bridge; S: Shell; D: Dust feature;  
2N: Double Nucleus; 3N: Triple Nucleus; A: Amorphous Halo; I: Irregular feature; and J: Jet. Brackets indicate uncertain
identification of the feature.}
\label{Bgq_control_individual}
\end{table*}

\begin{table*}
\centering
\begin{tabular}{lcccccccccl}
\hline
\hline
irac ID      & z$_{PHOT}$ &  N$_g$ & N$_b$ &  B$_{gq}$ &  N$_b^{av}$   & $\sigma$  &   B$_{gq}^{av}\pm\Delta B_{gq}^{av}$ & N$_b^{med}$ &  B$_{gq}^{med}$ & Morphology	  \\
 (1)         &  (2)       & (3)    & (4)   & (5)       & (6)           &   (7)     &  (8)	     &  (9)	   & (10)	     & (11)	 \\  
\hline
      006612 & 0.31  &       6    &    1.30  &     447  &    1.42  &	0.08   &    436$\pm$252   &	   1.45   &    433	& B,F,[D],[T   \\
      006613 & 0.30  &       6    &    1.26  &     446  &    1.37  &	0.08   &    436$\pm$248   &	   1.42   &    431	& B	       \\
      060958 & 0.40  &       3    &    1.75  &     132  &    1.74  &	0.08   &    133$\pm$216   &	   1.75   &    132	& T,[A],[B]    \\
      066504 & 0.39  &       0    &    1.71  &    -178  &    1.71  &	0.07   &   -178$\pm$113   &	   1.71   &   -178	& ...	       \\
      067417 & 0.39  &       1    &    1.49  &     -47  &    1.49  &	0.07   &    -47$\pm$137   &	   1.49   &    -48	& ...	       \\
      069266 & 0.35  &       5    &    1.79  &     344  &    1.77  &	0.07   &    345$\pm$266   &	   1.79   &    344	& ...	       \\
      072533 & 0.33  &       1    &    1.58  &     -58  &    1.57  &	0.08   &    -57$\pm$144   &	   1.58   &    -58	& S	       \\
      073242 & 0.41  &       3    &    1.67  &     137  &    1.68  &	0.07   &    137$\pm$212   &	   1.69   &    136	& ...	       \\
      074924 & 0.41  &       2    &    1.55  &      44  &    1.57  &	0.08   &     42$\pm$172   &	   1.58   &	41	& ...	       \\
      077695 & 0.35  &       2    &    1.70  &      31  &    1.71  &	0.07   &     30$\pm$184   &	   1.71   &	30	& T	       \\
      090430 & 0.38  &       2    &    1.75  &      27  &    1.77  &	0.07   &     24$\pm$187   &	   1.79   &	22	& A,F,[B]      \\
      092765 & 0.35  &       3    &    1.67  &     137  &    1.68  &	0.07   &    137$\pm$212   &	   1.69   &    136	& [A],[T]      \\
 093764$_-$1 & 0.39  &       1    &    1.67  &     -69  &    1.68  &	0.07   &    -70$\pm$151   &	   1.69   &    -71	& [S]	       \\
      094231 & 0.41  &       2    &    1.47  &      51  &    1.45  &	0.09   &     52$\pm$164   &	   1.47   &	51	& 2F,[T]       \\
      095727 & 0.38  &       2    &    1.55  &      44  &    1.57  &	0.08   &     42$\pm$172   &	   1.58   &	41	& F,S	       \\
      096307 & 0.34  &       2    &    1.45  &      52  &    1.42  &	0.08   &     55$\pm$164   &	   1.45   &	52	& I	       \\
      103198 & 0.38  &       1    &    1.64  &     -65  &    1.63  &	0.06   &    -64$\pm$148   &	   1.65   &    -66	& 2N,F,S       \\
      111427 & 0.32  &       4    &    1.52  &     244  &    1.52  &	0.07   &    244$\pm$221   &	   1.53   &    243	& 2N,T,2I      \\
      115327 & 0.35  &       4    &    1.52  &     244  &    1.52  &	0.07   &    244$\pm$221   &	   1.53   &    243	& 2F,[T]       \\
      115594 & 0.31  &       1    &    1.40  &     -39  &    1.49  &	0.07   &    -47$\pm$137   &	   1.49   &    -48	& 2N,T         \\
      118942 & 0.37  &       2    &    1.66  &      36  &    1.74  &	0.08   &     27$\pm$186   &	   1.75   &	26	& ...	       \\
      124509 & 0.34  &       5    &    1.26  &     352  &    1.37  &	0.08   &    342$\pm$230   &	   1.42   &    336	& B,2T,F       \\
      128074 & 0.34  &       3    &    1.55  &     148  &    1.63  &	0.06   &    139$\pm$206   &	   1.65   &    138	& B,[F]        \\
      132682 & 0.33  &       3    &    1.35  &     159  &    1.45  &	0.09   &    149$\pm$192   &	   1.47   &    147	& ...	       \\
      135859 & 0.40  &       3    &    1.26  &     164  &    1.37  &	0.08   &    153$\pm$186   &	   1.42   &    148	& [I]	       \\
      136904 & 0.39  &       1    &    1.70  &     -75  &    1.77  &	0.07   &    -83$\pm$161   &	   1.79   &    -84	& ...	       \\
      140456 & 0.30  &       5    &    1.43  &     352  &    1.52  &	0.07   &    343$\pm$242   &	   1.53   &    342	& 2T	       \\
      143149 & 0.37  &       4    &    1.43  &     253  &    1.52  &	0.07   &    244$\pm$221   &	   1.53   &    243	& T	       \\
      145098 & 0.32  &       1    &    1.51  &     -51  &    1.60  &	0.07   &    -61$\pm$147   &	   1.62   &    -62	& A,T	       \\
      147147 & 0.38  &       0    &    1.35  &    -129  &    1.45  &	0.09   &   -140$\pm$97    &	   1.47   &   -141	& [T]	       \\
      156161 & 0.30  &       4    &    1.70  &     235  &    1.63  &	0.06   &    242$\pm$231   &	   1.65   &    240	& T	       \\
      159936 & 0.41  &       6    &    1.75  &     439  &    1.68  &	0.07   &    447$\pm$277   &	   1.69   &    446	& 2N	       \\
      160500 & 0.34  &       1    &    1.50  &     -47  &    1.42  &	0.08   &    -39$\pm$131   &	   1.45   &    -42	& B,2T         \\   
      161724 & 0.34  &       0    &    1.69  &    -170  &    1.60  &	0.07   &   -162$\pm$106   &	   1.62   &   -163	& [F]	       \\
      166730 & 0.36  &       6    &    1.69  &     436  &    1.60  &	0.07   &    444$\pm$269   &	   1.62   &    443	& S,T	       \\
      173901 & 0.32  &       2    &    1.84  &      17  &    1.74  &	0.08   &     27$\pm$186   &	   1.75   &	26	& ...	       \\
      174667 & 0.34  &       4    &    1.70  &     235  &    1.63  &	0.06   &    242$\pm$231   &	   1.65   &    240	& [A]	       \\
      178868 & 0.37  &       0    &    1.75  &    -181  &    1.68  &	0.07   &   -173$\pm$111   &	   1.69   &   -174	& F	       \\
      181402 & 0.38  &       3    &    1.50  &     142  &    1.42  &	0.08   &    150$\pm$189   &	   1.45   &    147	& [I],[A]      \\
      181444 & 0.31  &       3    &    1.86  &     122  &    1.77  &	0.07   &    131$\pm$219   &	   1.79   &    129	& 2S,[I]       \\
      181914 & 0.36  &       0    &    1.43  &    -141  &    1.52  &	0.07   &   -150$\pm$101   &	   1.53   &   -150	& ...	       \\
      183081 & 0.36  &       2    &    1.63  &      39  &    1.71  &	0.07   &     30$\pm$184   &	   1.71   &	30	& F,[T]        \\
      184541 & 0.31  &       3    &    1.52  &     145  &    1.52  &	0.07   &    145$\pm$197   &	   1.53   &    145	& T	       \\
      186114 & 0.41  &       1    &    1.79  &     -84  &    1.77  &	0.07   &    -83$\pm$161   &	   1.79   &    -84	& T	       \\
      193735 & 0.37  &       0    &    1.58  &    -158  &    1.57  &	0.08   &   -157$\pm$104   &	   1.58   &   -158	& T	       \\
      193974 & 0.40  &       0    &    1.71  &    -178  &    1.71  &	0.07   &   -178$\pm$113   &	   1.71   &   -178	& [S]	       \\
      196827 & 0.37  &       4    &    1.55  &     235  &    1.45  &	0.09   &    245$\pm$215   &	   1.47   &    244	& T	       \\   
      198078 & 0.41  &       0    &    1.70  &    -173  &    1.63  &	0.06   &   -167$\pm$109   &	   1.65   &   -168	& [T]	       \\
      198996 & 0.32  &       3    &    1.86  &     122  &    1.77  &	0.07   &    131$\pm$219   &	   1.79   &    129	& ...	       \\
      203581 & 0.32  &       2    &    1.55  &      42  &    1.45  &	0.09   &     52$\pm$164   &	   1.47   &	51	& [T]	       \\
      207306 & 0.38  &       0    &    1.59  &    -165  &    1.68  &	0.07   &   -173$\pm$111   &	   1.69   &   -174	& ...	       \\
\hline		     			      		 
\end{tabular}						 
\caption{Same as in Table \ref{Bgq_individual} but for the 51 early-type galaxies in the EGS* sample.}
\label{Bgq_control_individual2}
\end{table*}

\section*{Acknowledgments}

CRA ackowledges financial support from the Spanish Ministry of Science and Innovation (MICINN) through project 
Consolider-Ingenio 2010 Program grant CSD2006-00070: First Science with the GTC 
(http://www.iac.es/consolider-ingenio-gtc/), the Estallidos group through project PN AYA2010-21887-C04.04 and 
STFC PDRA (ST/G001758/1). PSB acknowledges support in the form of an STFC PhD studentship. 
KJI is supported through the Emmy Noether programme of the German Science Foundation (DFG).

This research has made use of the NASA/IPAC Extragalactic Database (NED) and the NASA/ IPAC Infrared Science Archive,
which are operated by the Jet Propulsion Laboratory, California Institute of Technology, under 
contract with the National Aeronautics and Space Administration.
GAIA was created by the now closed Starlink UK project, funded by the Particle Physics and Astronomy 
Research Council (PPARC) and has been more recently supported by the Joint Astronomy Centre Hawaii funded 
again by PPARC and more recently by its successor organisation the Science and Technology Facilities Council (STFC). 

We finally acknowledge thorough and useful comments from the anonymous referee.

\label{lastpage}

\end{document}